\documentclass[conference]{IEEEtran}

\makeatletter
\def\endthebibliography{%
  \def\@noitemerr{\@latex@warning{Empty `thebibliography' environment}}%
  \endlist
}
\makeatother

\usepackage{cite} 
\PassOptionsToPackage{hyphens}{url}\usepackage[hyperfootnotes=false,bookmarks=false]{hyperref} 
\usepackage{xspace} 
\usepackage{mwe}
\usepackage{soul}
\usepackage{graphicx}
\usepackage{amsmath}
\usepackage{amssymb}
\usepackage{arydshln}
\usepackage{subcaption}
\usepackage{multicol}
\usepackage{multirow}
\usepackage{makecell}
\usepackage{booktabs}
\usepackage{threeparttable}
\usepackage{wasysym}
\usepackage[inline]{enumitem}
\usepackage{textgreek}
\usepackage{upgreek}
\usepackage{pifont}
\usepackage{array}
\usepackage{colortbl}
\usepackage{scalerel}
\usepackage[dvipsnames]{xcolor}
\usepackage{tikz}
\usepackage{tcolorbox}
\usepackage[font=small,labelfont=bf]{caption}
\usepackage{setspace}

\pdftrailerid{} 
\usepackage[inline,nomargin,index]{fixme}

\FXRegisterAuthor{kav}{akav}{\textcolor{orange}{Kav}}

\FXRegisterAuthor{pit}{apit}{\textcolor{red}{PIT}}
\newcommand{\pit}[1]{\pitnote{\textcolor{red}{#1}}}

\FXRegisterAuthor{cri}{acri}{\textcolor{blue}{Cri}}

\FXRegisterAuthor{ema}{aema}{\textcolor{LimeGreen}{Ema}}
\newcommand{\ema}[1]{\emanote{\textcolor{LimeGreen}{#1}}}

\FXRegisterAuthor{hjb}{ahjb}{\textcolor{green}{HJB}}

\newcommand{\tocheck}[1]{{#1}}
\newcommand{\hhd}[1]{{#1}}
\newcommand{\hh}[1]{{#1}}
\newcommand{\pp}[1]{{#1}}
\newcommand{\ee}[1]{{#1}}
\newcommand{\hht}[1]{{#1}}
\newcommand{\hhf}[1]{{#1}}
\newcommand{\hhn}[1]{{#1}}
\newcommand{\hhc}[1]{{#1}}
\newcommand{\pph}[1]{{#1}}
 \newcommand{\eeh}[1]{{#1}}

\usepackage{listings}
\usepackage{lstautogobble}
\usepackage{expl3,xparse}
\ExplSyntaxOn
\NewDocumentCommand \lstcolorlines { O{gray!30} m }
{
    \clist_if_in:nVT { #2 } { \the\value{lstnumber} }{ \color{#1} }
}
\ExplSyntaxOff

\definecolor{plainText}{RGB}{74,131,31}
\definecolor{comments}{RGB}{74,131,31}
\definecolor{strings}{RGB}{180,54,34}
\definecolor{numbers}{RGB}{44,45,211}
\definecolor{keywords}{RGB}{160,49,158}
\definecolor{preprocessorStatements}{RGB}{109,75,48}
\definecolor{classNames}{RGB}{101,63,165}

\usepackage{float}
\newfloat{lstfloat}{tbhp}{lop}
\floatname{lstfloat}{Listing}
\usepackage{dblfloatfix}

\newcommand{\circled}[1]{\raise.17ex\hbox{{\textcircled{\scriptsize{#1}}}}}
\newcommand\hmm[1]{\ifnum\ifhmode\spacefactor\else2000\fi>1000 \uppercase{#1}\else#1\fi}

\newcommand{\n}{\emph{TRRespass}\xspace}
\newcommand{\rh}{RowHammer\xspace}
\newcommand{\js}{JavaScript\xspace}

\newcommand{\tildesmall}{{\raise.17ex\hbox{$\scriptstyle\sim $}}}

\newcommand{\trefi}{{\ttfamily tREFI}\xspace}
\newcommand{\trfc}{{\ttfamily tRFC}\xspace}

\newcommand{\tmaw}{{\ttfamily tMAW}\xspace}
\newcommand{\mac}{{\ttfamily MAC}\xspace}
\newcommand{\nDimms}{{42}\xspace}

\newcommand{\nVulnDimms}{{13}\xspace}
\newcommand{\activate}{{\ttfamily ACTIVATE}\xspace}
\newcommand{\act}{{\ttfamily ACT}\xspace}
\newcommand{\precharge}{{\ttfamily PRECHARGE}\xspace}

\newcommand{\refresh}{{\ttfamily REFRESH}\xspace}

\newcommand{\yes}{{\ttfamily \color{ForestGreen}\checkmark}}%
\newcommand{\notavail}{{\ttfamily \rule[.5ex]{.75em}{0.25pt}}}%
\DeclareRobustCommand{\ulcolored}[2]{{\setulcolor{#1}\ul{#2}}}
\newcommand{\stripe}{\rowcolor{blue!5}}
\newcommand{\mr}[2]{\multicolumn{1}{c}{\multirow{#1}{*}{\makecell{#2}}}}
\newcommand{\mcr}[1]{$\mathcal{B}_{\fontsize{9}{4}\selectfont #1}$}
\newcommand{\smg}[1]{$\mathcal{A}_{\fontsize{9}{4}\selectfont #1}$}
\newcommand{\hnx}[1]{$\mathcal{C}_{\fontsize{9}{4}\selectfont #1}$}
\newcommand{\nsided}[1]{#1-sided}
\newcommand{\dnsided}[2]{$\langle$#2-sided $\mid$ dist=#1$\rangle$}
\newcommand{\manysided}{many-sided\xspace}
\newcommand{\Manysided}{Many-sided\xspace}
\newcommand{\observation}[1]{\textbf{\emph{Observation\,}#1:}\xspace}
\newcommand{\aarch}{ARMv8\xspace}
\newcommand{\nPhones}{13\xspace}
\newcommand{\nVulnPhones}{5\xspace}
\newcommand{\lpddr}{LPDDR4(X)\xspace}
\newcommand{\nHynixVuln}{2\xspace}
\newcommand{\nHynixTot}{14\xspace}

\begin{document}

\title{TRRespass: Exploiting the Many Sides of\\ Target Row Refresh}

\newcommand\blfootnote[1]{%
  \begingroup
  \renewcommand\thefootnote{}\footnote{#1}%
  \addtocounter{footnote}{-1}%
  \endgroup
}

\newcommand*{\affaddr}[1]{#1} 

\makeatletter
\newcommand{\linebreakand}{%
  \end{@IEEEauthorhalign}
  \hfill\mbox{}\par
  \mbox{}\hfill\begin{@IEEEauthorhalign}
}
\makeatother

 \author {
       \IEEEauthorblockN{
         Pietro Frigo\IEEEauthorrefmark{1}\IEEEauthorrefmark{2}\hspace{4mm}
         Emanuele Vannacci\IEEEauthorrefmark{1}\IEEEauthorrefmark{2}\hspace{4mm}
         Hasan Hassan\IEEEauthorrefmark{4}\hspace{4mm}
         Victor van der Veen\IEEEauthorrefmark{5}\\
         Onur Mutlu\IEEEauthorrefmark{4}\hspace{4mm}
         Cristiano Giuffrida\IEEEauthorrefmark{1}\hspace{4mm}
         Herbert Bos\IEEEauthorrefmark{1}\hspace{4mm}
         Kaveh Razavi\IEEEauthorrefmark{1}}

     \linebreakand
     \IEEEauthorblockA{
             \IEEEauthorrefmark{1}Vrije Universiteit Amsterdam
     }
     \and
     \IEEEauthorblockA{
         \IEEEauthorrefmark{4}ETH Z\"urich
     }
     \and
     \IEEEauthorblockA{
         \IEEEauthorrefmark{5}Qualcomm Technologies Inc.
     }

     \linebreakand
     \IEEEauthorblockA{
     \small{\IEEEauthorrefmark{2}Equal contribution joint first authors}
     }
     }

\maketitle

\maketitle
\pagestyle{plain}

\setstretch{.93}

\begin{abstract}
After a plethora of high-profile \rh attacks, CPU and DRAM \hhf{vendors}
scrambled to deliver what was meant to be the definitive hardware solution
against the \rh problem: \emph{Target Row Refresh} (\emph{TRR}). A common belief
among practitioners is that, \hhf{for} the latest generation of DDR4 systems that are
protected by TRR, \rh is no longer an issue in practice. However, in
reality, very little is known about TRR. 
\hhf{How does TRR exactly prevent \rh? Which parts of a system are
responsible for operating the TRR mechanism? Does TRR completely solve the \rh
problem or does it have weaknesses?}
%

In this paper, we demystify the inner workings of TRR and debunk its
security guarantees. We show that what is advertised as a single
mitigation \hhf{mechanism} is actually a series of different solutions coalesced under the
umbrella term Target Row Refresh. We inspect and disclose, \hht{via} a deep
analysis, different \hht{existing} TRR solutions and demonstrate that modern
implementations operate entirely inside DRAM chips. Despite the
difficulties of analyzing \hht{in-DRAM} mitigations, we describe novel
techniques for gaining insights into the operation of these \hhf{mitigation
mechanisms}. These insights \hht{allow} us to build \n, a scalable
black-box \rh fuzzer that we \hht{evaluate} on \nDimms recent DDR4
\hhn{modules}. 

\n shows that even the
\hh{latest generation} DDR4 \hhf{chips} with \hht{in-DRAM} TRR, immune to all known
\rh attacks, are often \emph{still vulnerable} to \hht{new} TRR-aware variants \hh{of
\rh} \hht{that we develop}. In particular, \n finds that, on present-day
\hht{DDR4 modules,} \rh is still
possible when \emph{many} aggressor rows are used (\hhn{as many as} 19 in some cases),
\hhn{with a method} we generally refer to as \emph{\Manysided \rh}. Overall, our
analysis shows that \nVulnDimms out of the \nDimms{} \hhn{modules} from
all three major DRAM \hhf{vendors} (\hhf{i.e.}, Samsung, Micron\hhc{,} and Hynix) are vulnerable to our TRR-aware \hht{\rh}
access patterns\hht{, and thus one can still mount existing
state-of-the-art \hhf{system-level} \rh attacks}. \hh{In addition to DDR4, we also experiment with
\lpddr\footnote{\hhf{We refer to both LPDDR4 and LPDDR4X chips as
LPDDR4(X).}} chips and show that they} are susceptible to
\rh bit flips too. Our results provide concrete evidence that the pursuit of
better \hhn{\rh} mitigations must continue.

\vspace{-2mm}
    \blfootnote{
    \IEEEauthorrefmark{5}Victor contributed to the research on DDR4 modules.
    }
\end{abstract}


\section{Introduction}
\label{sec:intro}

Is \rh a solved problem? The leading DRAM vendors have already answered
this question with a resounding ``yes'', advertising \hh{the latest
generation} DDR4 systems as \rh-free and \hh{using} \emph{Target Row
Refresh} (\emph{TRR}) as the \hht{``silver bullet''} that \hh{eradicates} the
vulnerability~\cite{micron2016ddr4,samsung2014trrDram}. Unfortunately, very
little is known about the actual implementation or security of TRR on
modern systems. \hh{Even the major consumers of DRAM in the industry} have to simply
take the DRAM vendors at their word \hh{as the vendors do not disclose
the details of the TRR schemes they implement}. In this paper, we question this
\emph{security by obscurity} strategy and analyze the mechanisms
behind TRR to bypass this prevalent mitigation.
Our results are
worrisome, showing that \rh is not only still unsolved, but also that the
vulnerability is widespread \hh{even in} \hht{latest off-the-shelf} DRAM chips. Moreover, once
the \rh mitigation \hhf{mechanism} is \hhf{turned off,\footnote{We turn off the in-DRAM \rh mitigation
mechanism by disabling \refresh commands\hhn{,} as we explain in
Section~\ref{sec:rev-dram}.}} we observe bit flips with \hht{as few} as
\hhn{45}K
\hh{DRAM row activations}, showing \hh{that} DDR4 \hht{and \lpddr} chips are more vulnerable
\hh{to \rh} than their DDR3 predecessors\hh{, which \hht{can tolerate} much higher
row activation counts \hhf{(\hhn{e.g.},
\tildesmall{}139K~\cite{kim2014flipping})}}.

\vspace{2mm}
\noindent\textbf{RowHammer.}
\hhn{Within} \hht{only} five years since its discovery, exploits based on the \rh
vulnerability~\cite{kim2014flipping} have spread to almost every type
of computing
system\hht{~\cite{mutlu2019rowhammer, mutlu2017rowhammer}}. Personal computers~\cite{seaborn2015exploiting-1,gruss2016rowhammer,bosman2016dedup,tatar2018defeating,gruss2018another},
cloud servers~\cite{razavi2016flip,xiao2016one,cojocar2019eccploit,tatar2018throwhammer,rambleed,poddebniak2018attacking,236248}, and mobile
phones~\cite{vanderveen2016drammer,vanderveen2018guardion,frigo2018grand} have all fallen
victim to attacks with \rh bit flips triggered from native
code~\cite{seaborn2015exploiting-1,razavi2016flip,xiao2016one,cojocar2019eccploit,qiao2016new,gruss2018another,bhattacharya2016curious,poddebniak2018attacking,236248}, JavaScript in the
browser~\cite{gruss2016rowhammer,bosman2016dedup,frigo2018grand,seaborn2015exploiting-1}
and even remote clients across the
network~\cite{tatar2018throwhammer,lipp2018nethammer}. From an academic
\hht{demonstration}, the \rh vulnerability \hht{has} evolved into a major
\hht{security vulnerability} for the entire industry.
In response,
hardware vendors have scrambled to address the \rh issue. 

\vspace{2mm}
\noindent\textbf{Target Row Refresh.}
Reliable solutions against \rh simply do not exist for older hardware and
stopgap solutions such as \hh{using ECC} and doubling \hht{(or even
quadrupling)} the refresh rate have
proven ineffective~\cite{aweke2016anvil,cojocar2019eccploit,
kim2014flipping}. \hhn{In} the
early days of the DDR4 specification, DRAM vendors announced they would
deploy \hht{the} \hh{Target Row Refresh (TRR)} \hht{mitigation}
\hhn{mechanism} on
newer-generation \hht{DDRx} systems to eradicate the \rh vulnerability~\cite{samsung2014trrDram,micron2016ddr4}. 
While \hhn{reports} of bit flips on DDR4 devices~\cite{lipp2018nethammer,thirdio_nodate_rowhammer,gruss2018another}
suggest that the deployment of \hhn{such} mitigation \hhn{mechanisms} may not
have been prompt, 
it is commonly assumed that  
TRR technology on recent DDR4 systems has put an end to \rh
attacks~\cite{oracleRambleed,arsRambleed}. \hhf{Nowadays,} the
leading DRAM vendors explicitly advertise \rh-free
modules~\cite{micron2016ddr4,samsung2014trrDram}. \hh{Our} initial
assessment \hhf{confirms} that none of the \emph{known} \rh variants produce bit flips on \nDimms recent DDR4 modules.
However, little is known about TRR beyond what its name suggests, namely that it generates extra refreshes for rows targeted by \rh.



%
\vspace{2mm}
\noindent\textbf{The many sides of TRR.}
In this paper, we take a closer look at the TRR implementations on modern
systems. In contrast to what the literature
suggests~\cite{lipp2018nethammer}, we show that
TRR is not a single mitigation \hhf{mechanism} but rather a \hhn{family} of
solutions, 
implemented either in the CPU's memory controller or in
the DRAM chips themselves. One of the best-known implementations of
TRR-like functionality, \hhn{Intel's} \emph{pTRR}~\cite{intel2014ptrr}, appeared in the memory controllers of
Intel CPUs as early as \hht{2014} to protect vulnerable DDR3
modules. Interestingly, while memory controller-based
\hhn{TRR} implementations still \hhn{exist in} modern DDR4 systems, we show that they are
now mostly dormant. This is presumably because such functionality is
considered superfluous now that the DRAM vendors advertise \rh-free
modules with \hhn{in-DRAM TRR, i.e.,} TRR implemented entirely inside the
\hhn{DRAM} chips~\cite{micron2016ddr4,samsung2014trrDram}.

\hh{Unfortunately}, none of the \hhf{in-DRAM} TRR variants are \hhf{well} documented. As a result, their
security guarantees are buried deep inside the \hhf{DRAM chips} \hht{that embed} them. This
poses a major threat to the security of modern systems, if they turn out to
be vulnerable after all. 

\vspace{2mm}
\noindent\textbf{\n.}
To compensate for the lack of information, we \hhf{investigate} the mechanisms behind
TRR and show that \hht{new} TRR-aware attacks can still exploit \hhf{the}
\rh vulnerability on modern DDR4
\hhf{devices}. We start our analysis by investigating TRR variants
implemented in the memory controller and DRAM chips. 
\hhf{We inspect memory controller-based TRR mechanisms using timing
side channels to analyze when the memory controller performs a targeted
refresh or whether it refreshes the entire DRAM at increased rate.
Inspecting more recent in-DRAM TRR mechanisms is more challenging since
these mechanisms operate transparently to the memory controller, and thus
the rest of the system (e.g., targeted refresh may \hhn{or may not} happen
during the fixed \trfc refresh latency).}
%
%
To
address this challenge, we \hh{use SoftMC~\cite{hassan2017softmc}, an}
FPGA-based memory controller. \hh{SoftMC} provides us with fine-grained
control over the commands sent to DRAM. Using \rh bit flips and a careful
selection of DRAM commands, we gradually reconstruct the different
mitigations deployed on \hh{recent DDR4} \hht{modules}, \hhf{and}
\hhn{uncover} how they track the rows being hammered and how they protect
\hh{the victim rows}.

Our analysis shows that, while TRR implementations differ across DRAM
vendors, most \hh{TRR variations} can be bypassed by what we \hht{introduce
as} \emph{\Manysided \rh} (i.e., \rh with many aggressor rows). Building on
this insight, we present \n
to identify TRR-aware \rh
access patterns on modern systems. Our fuzzing strategy generates \manysided \rh patterns in an entirely
black-box fashion, without relying on any implementation details of the memory controller or
\hh{DRAM} chips. We show \hh{that} relatively simple \manysided \rh patterns
identified by \n can successfully trigger bit flips on DDR4 DRAM chips
from all three \hht{major} DRAM vendors, namely Samsung, Micron\hhc{,} and Hynix (\hht{representing} over 95\% of the DRAM
market~\cite{dram-share})\pp{, as well as on \hhc{mobile} phones employing \lpddr DRAM chips}.
Overall, our analysis
provides evidence \hh{for} significant weaknesses in state-of-the-art TRR
implementations, showing they can be bypassed to expose the
vulnerable DDR4 substrate to state-of-the-art \hhf{system-level} \rh attacks.

\vspace{3mm}
\noindent\textbf{Contributions.}
We make the following contributions:
\begin{itemize}
    \item We present the first overview of different \hht{Target
        Row Refresh (TRR)} implementations available on modern
        systems\hht{, which have been publicized as an effective solution
        to the \rh problem.} 
    \item We \hht{analyze the} \hhf{memory-controller-based} and
        \hht{in-DRAM} TRR implementations by the leading hardware vendors.
	\item We present \n, a black-box \rh fuzzer, which can
        automatically identify TRR-bypassing \rh access patterns on
        \nVulnDimms of \nDimms tested DDR4 \hht{modules from all three major DRAM
        manufacturers} \hhn{as well as \nVulnPhones of \nPhones tested
        \hhc{mobile} phones}.
    \item We \hhf{use the} \rh access patterns \hhf{that \n identifies} on modern
        \hht{TRR-protected} DDR4 and LPDDR4(X) DRAM chips \hhf{to} show how attackers can use
        TRR-aware \hht{\rh access} patterns to mount state-of-the-art \rh attacks on these modules.
\end{itemize}



\section{\rh on DDR4: still a problem?}
\label{sec:bg}

\hhf{Prior} research has characterized~\cite{kim2014flipping,
tatar2018defeating, vanderveen2016drammer, cojocar2019eccploit} and
exploited \hh{the \rh vulnerability of DRAM}~\cite{razavi2016flip,
frigo2018grand, gruss2018another, gruss2016rowhammer, tatar2018throwhammer,
cojocar2019eccploit, xiao2016one}. While there has been \hh{systematic}
research on the vulnerability \hh{on} DDR3 systems~\cite{kim2014flipping,
tatar2018defeating}, \hhn{relatively} little is known about the extent of \rh on
\hht{recent} DDR4 systems. In this section, we \hh{first} provide
\hh{the necessary background on DRAM and \rh for understanding the rest of
the paper}. \hh{We refer the reader to prior work~\hhn{\cite{kim-isca2012,
kim2014flipping, zhang2014half, hassan2016chargecache, lee-hpca2013,
seshadri2017ambit, chang2017understanding, chang2016understanding,
chang2014improving, chang2016low, lee-hpca2015, lee2017design,
lee2015decoupled, liu2013experimental, liu2012raidr, seshadri2013rowclone,
seshadri2015gather, hassan2019crow, lee2016simultaneous, seshadri2019dram}} for a more detailed description of DRAM
organization and operation.} \hh{Then, we} perform a preliminary analysis
on recent DDR4 systems using existing ``hammering'' patterns in the
literature~\cite{kim2014flipping,tatar2018defeating,gruss2018another} to
investigate the current status of the \rh vulnerability \hht{on DDR4}.


\subsection{DRAM Organization}
\label{sec:bg:dram}

Figure~\ref{fig:dram:high} depicts \hht{the} high-level organization of
\hh{a DRAM-based main memory subsystem}. The CPU communicates with \hh{DRAM}
through the \emph{Memory Controller} (from now on also referred to as MC).
\hhf{The} \hh{MC is responsible for} \hhf{issuing} memory requests to
\hh{the corresponding DRAM \emph{channel}. DRAM channels operate independently from
each other and a single channel can host multiple memory modules (or
\emph{DIMMs}). DRAM \emph{chips} in a DIMM are organized as a single
\hht{\emph{rank}} or multiple \emph{ranks}. The DRAM chips that form a rank
operate in lock-step, simultaneously receiving the same DRAM command but
operating on different data portions. Thus, a rank composed of several DRAM
chips appears as a single large memory to the system. A DRAM chip contains
multiple DRAM \emph{banks} that operate in parallel.}

\begin{figure}[!ht]
	        \centering
	        \includegraphics[width=\linewidth]{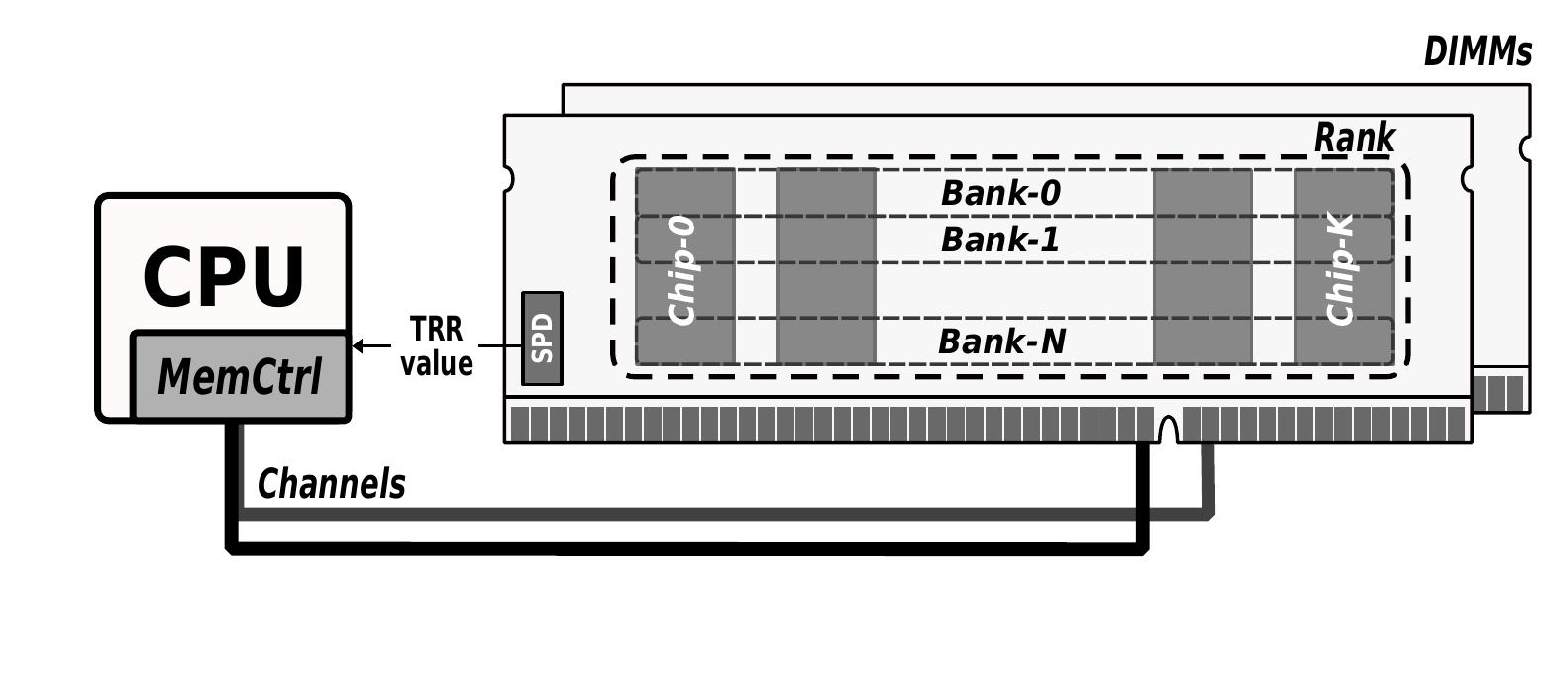}
	        \vspace{-10mm}
	        \caption{High-level DRAM organization.}
	        \label{fig:dram:high}
\end{figure}

\vspace{2mm}\noindent\textbf{Inside a bank.}
A \hhf{DRAM} bank can be logically seen as a \hht{two-}dimensional array of \hh{DRAM}
\emph{cells} (Figure~\ref{fig:dram:low}). \hh{Cells that share a
\emph{wordline} are referred to as a DRAM \emph{row}. The \emph{row
decoder} selects (i.e., activates) a row to load its data into the \emph{row buffer}, where
data can be read and modified.}
A \hh{DRAM} cell consists of two
\hh{components}: 
\begin{enumerate*}[label=(\roman*)]
    \item a \emph{capacitor} and
    \item \hh{an \emph{access transistor}}
\end{enumerate*}.
The capacitor \hh{stores} a single bit of information \hh{as electrical
charge}. \hh{During an access to a cell, the corresponding wordline enables
the access transistor of the cell, which connects the cell capacitor to the
\emph{bitline}}. \hh{Thus, to read/write data in a specific DRAM row, the
memory controller first issues an \activate command to bring the row's data
into the row buffer. The row buffer consists of \emph{sense
amplifiers}, each connected to a bitline. Because row activation destroys the data stored in
\hht{the cell capacitor}, a sense amplifier not only successfully determines the bit
stored in the cell, but also restores the charge back \hht{into} the capacitor.
After \hht{the activated row of cells} is fully restored, the memory
controller \hht{can issue} a \precharge
command to close the row and prepare the bank for activating a different
row.}


\begin{figure}[!ht]
	        \centering
	        \includegraphics[width=0.8\linewidth]{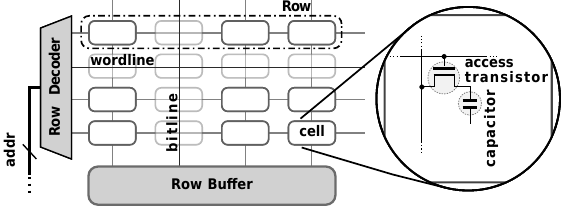}
            \caption{\hht{DRAM bank} organization \hht{(logical)}.}
	        \label{fig:dram:low}
\end{figure}

DRAM cell capacitors \hhf{are not ideal and they} \hh{gradually
lose their charge} over time. Thus, the memory controller needs to
\emph{refresh} the contents of \hht{all} cells \hh{periodically} (usually
\hh{every} $64\,ms$~\cite{jedec2014lpddr4, jedec2014ddr4, liu2012raidr}) to prevent data loss. 

\subsection{\rh}
\label{sec:bg:rh}
\rh is a well-known DRAM vulnerability that has been investigated since
2012~\cite{kim2014flipping,intelcorp2016row,intelcorp2016distributed,intelcorp2015method,intelcorp2014method}.
\hh{When a particular DRAM row is repeatedly activated and precharged many
times (i.e., hammered), electro-magnetic interference between the hammered row and its
neighbor rows can cause the cell capacitors in the neighbor rows to leak
much faster \hht{than under} normal operation. Rows that are hammered are
referred to as \emph{aggressor} rows, whereas \hht{their} neighbor rows are
referred to as \emph{victim} rows.} Kim et al.~\cite{kim2014flipping}
\hh{are} the first to perform a large-scale study of the properties of
\rh bit flips on DDR3 modules. They \hh{report} \tildesmall
85\% of the tested modules to be vulnerable to \rh. Since \hh{one can cause
\rh bit flips \hht{solely by} performing memory accesses}, \rh quickly became \hh{a
popular vector} for \hh{developing} real-world
attacks~\hht{\cite{seaborn2015exploiting-1, gruss2016rowhammer,
vanderveen2016drammer, tatar2018throwhammer, cojocar2019eccploit,
frigo2018grand, xiao2016one, razavi2016flip, gruss2018another,
tatar2018defeating, vanderveen2018guardion, lipp2018nethammer,
pessl2016drama, aga2017good, barenghi2018software, bhattacharya2016curious,
bhattacharya2018advanced, carre2018openssl, fournaris2017exploiting,
jang2017sgx, poddebniak2018attacking, qiao2016new, zhang2018triggering}}.

\vspace{3mm}
\noindent\textbf{Attacks}. Seaborn and
Dullien~\cite{seaborn2015exploiting-1} initially demonstrated \rh attacks
\hh{for compromising} the Linux kernel. Afterwards, other researchers
exploited \rh to break cloud isolation~\cite{razavi2016flip,xiao2016one,cojocar2019eccploit,tatar2018throwhammer,rambleed,poddebniak2018attacking,236248}, ``root'' mobile
devices~\cite{vanderveen2016drammer,vanderveen2018guardion}, \hht{take
over} browsers~\cite{gruss2016rowhammer,bosman2016dedup,frigo2018grand},
and \hht{attack} server applications over the
network~\cite{tatar2018throwhammer,lipp2018nethammer}. All these attacks
\hh{demonstrate} the severity of the \hh{\rh} threat and the need to build
effective defenses.

\vspace{3mm}
\noindent\textbf{Defenses}. \hh{Various software-based \rh} defenses
advocate for the detection of \hh{the} \rh patterns~\cite{aweke2016anvil},
cross-domain~\cite{brasser2017catt} \hh{(}or more general\hh{)} memory
isolation~\cite{vanderveen2018guardion, tatar2018throwhammer,
konoth2018zebram}, or software-controlled ECC~\cite{cojocar2019eccploit}.
Unfortunately, these defenses are complex, expensive, and/or incomplete.
As a result, they are not deployed in practice. \hht{Immediately-deployable} hardware-based defenses,
such as doubling \hht{(or even quadrupling)} the refresh rate or
\hh{\hht{using} existing DRAM modules
with error-correction \hht{code} (ECC) capability to protect against \rh},
\hht{are used in the field}, \hht{yet they} have been shown to be
insecure~\cite{cojocar2019eccploit, aweke2016anvil, kim2014flipping}.

\vspace{3mm}
\noindent\textbf{DDR4: Towards a \rh-less landscape}.
\pph{ Most \hh{prior} \rh research \hh{focuses} on DDR3
	systems~\hht{\cite{kim2014flipping, seaborn2015exploiting-1,
			gruss2016rowhammer, vanderveen2016drammer, tatar2018throwhammer,
			cojocar2019eccploit, frigo2018grand, xiao2016one, razavi2016flip,
			tatar2018defeating, vanderveen2018guardion, aga2017good,
			barenghi2018software, bhattacharya2016curious,
			bhattacharya2018advanced, carre2018openssl, fournaris2017exploiting,
	jang2017sgx, poddebniak2018attacking, qiao2016new, zhang2018triggering}}.
	While there are reports of bit flips on DDR4 chips in prior
	work~\cite{lipp2018nethammer,gruss2018another,thirdio_nodate_rowhammer},
	these results are on earlier generations of DDR4. Through communication with
	industry, we have confirmation that some early-generation DDR4 chips did not
	have the ``TRR'' mitigation enabled by default. In order to understand the
	modern landscape we test a set of \nDimms recent DDR4 modules against all
	standard hammering patterns: \begin{enumerate*}[label=(\roman*)] \item
		\emph{single-sided}, which simply activates two arbitrary
		(\emph{aggressor}) rows in the same bank to induce bit flips in
		\hh{their} adjacent \emph{victim} rows (Figure~\ref{fig:single-sided});
	\item \emph{double-sided}, which uses the same access patterns as
        \emph{single-sided} but the two aggressor rows are chosen to
            \hhn{both be adjacent to} a
		single victim row to amplify the effect \hh{of hammering}
		(Figure~\ref{fig:double-sided}); and \item \emph{one-location}, which
		activates a single row (Figure~\ref{fig:one-location}) and only applies
		to systems where the MC employs a
		\hh{closed}-row~\cite{intelcorp2016closePg, kim2010atlas} or
		adaptive~\cite{intelcorp2004adaptive} page policy. \end{enumerate*}
		\begin{figure}[!ht] 
\centering
\begin{subfigure}[t]{.3\linewidth}
	\centering\includegraphics[width=1\linewidth]{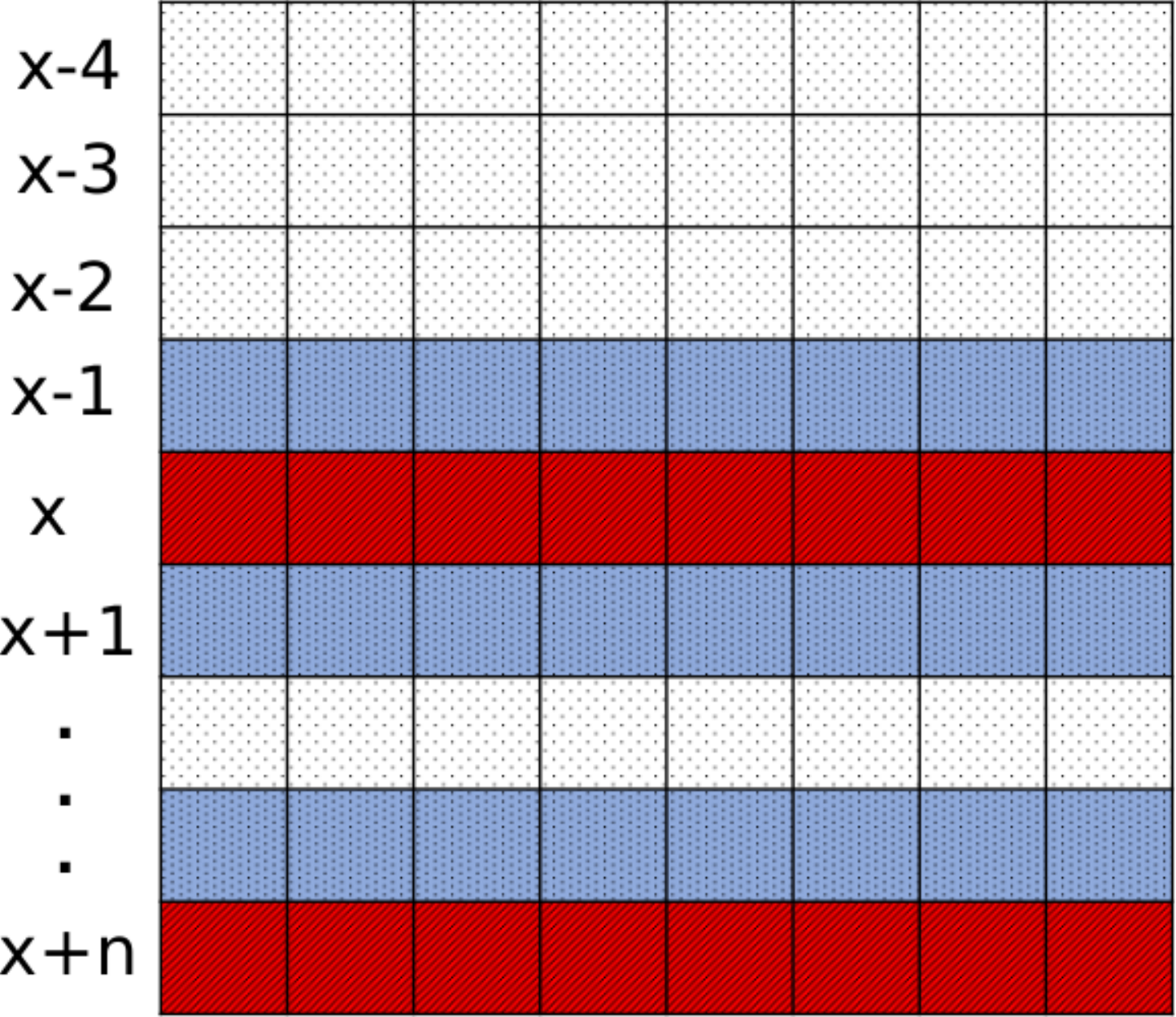}
	\caption{Single-sided} \label{fig:single-sided} 
\end{subfigure} \hfill
\begin{subfigure}[t]{.3\linewidth}
	\centering\includegraphics[width=1\linewidth]{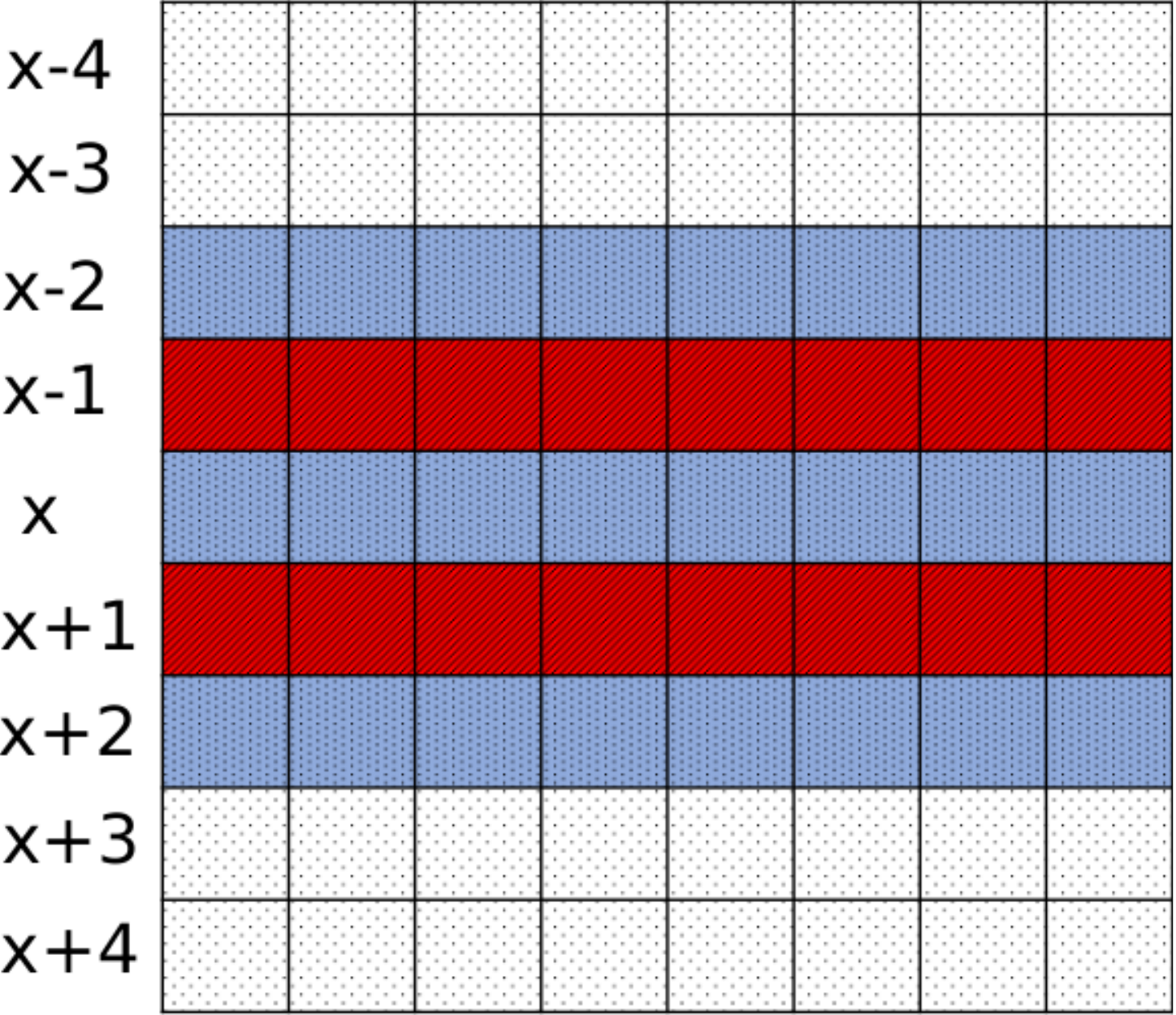}
	\caption{Double-sided} \label{fig:double-sided} 
\end{subfigure} \hfill
\begin{subfigure}[t]{.3\linewidth}
	\centering\includegraphics[width=1\linewidth]{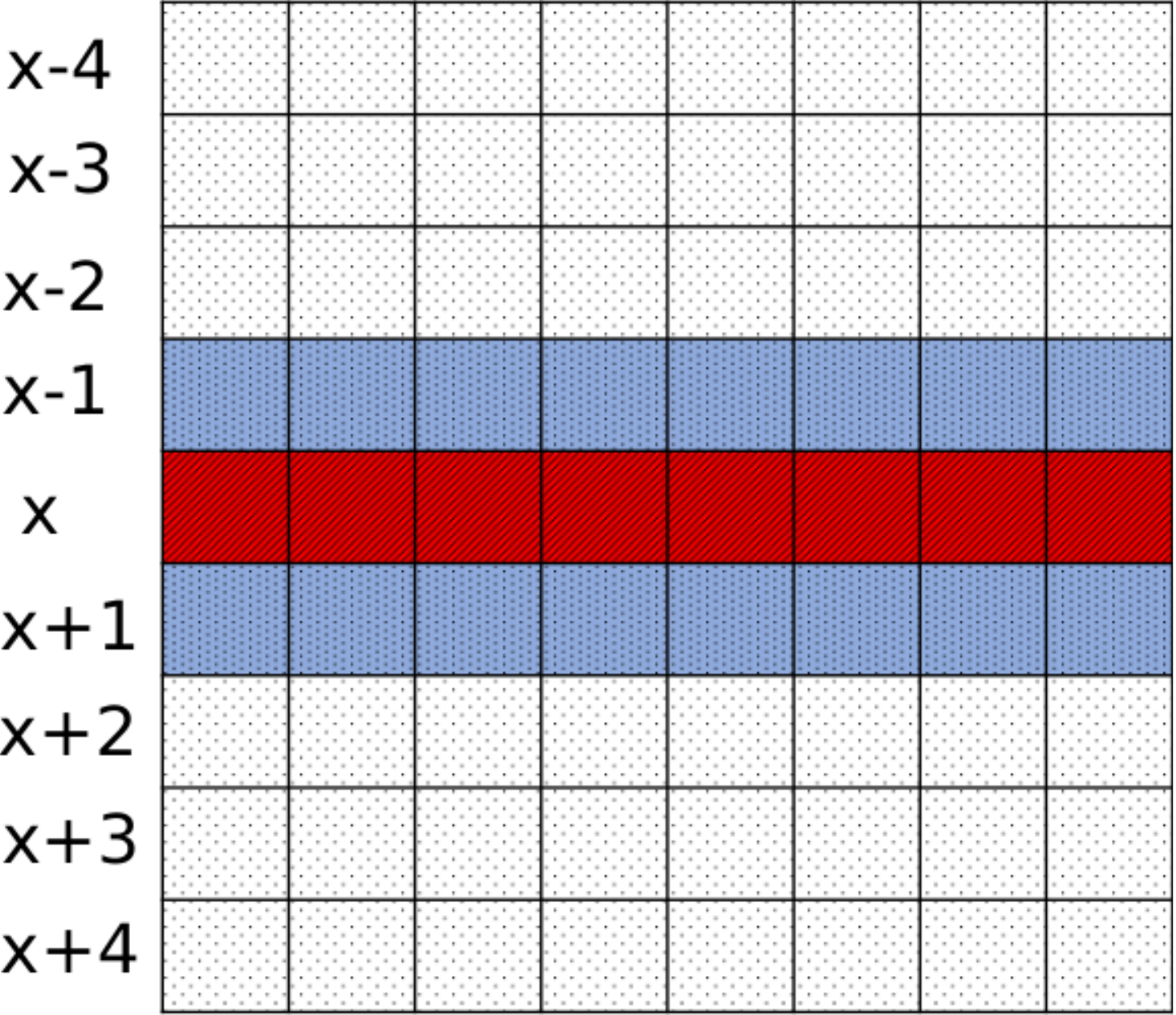}
	\caption{One-location} \label{fig:one-location} 
\end{subfigure} \hfill
    \caption{Standard hammering patterns. \hh{The aggressor} rows are
    \hh{highlighted} in red (\protect\tikz \fill[red] (0.1,0.1) rectangle
    (0.3,0.3);) and victim rows are \hh{highlighted} in blue (\protect\tikz \fill[blue] (0.1,0.1) rectangle (0.3,0.3);).}
\end{figure}

 As \hh{we show} in
		Figure~\ref{fig:rh-effectiveness}, our analysis \hh{reveals} that none
		of these patterns manifest any \hh{bit flip} on \hh{the modules we
		test}, even when using the \hh{exact} test suites provided by prior
		work~\cite {gruss2018another,tatar2018defeating, kim2014flipping}.
    \hhn{Our} results suggest that recent
	DDR4 \hh{chips} include effective mitigations against the best \emph{known}
	hammering patterns, matching claims of DRAM
	vendors~\cite{micron2016ddr4,samsung2014trrDram}. This raises the important
question: \emph{Is \rh a solved problem?} }

\begin{figure}[t]        
	\centering
	\includegraphics[width=\linewidth]{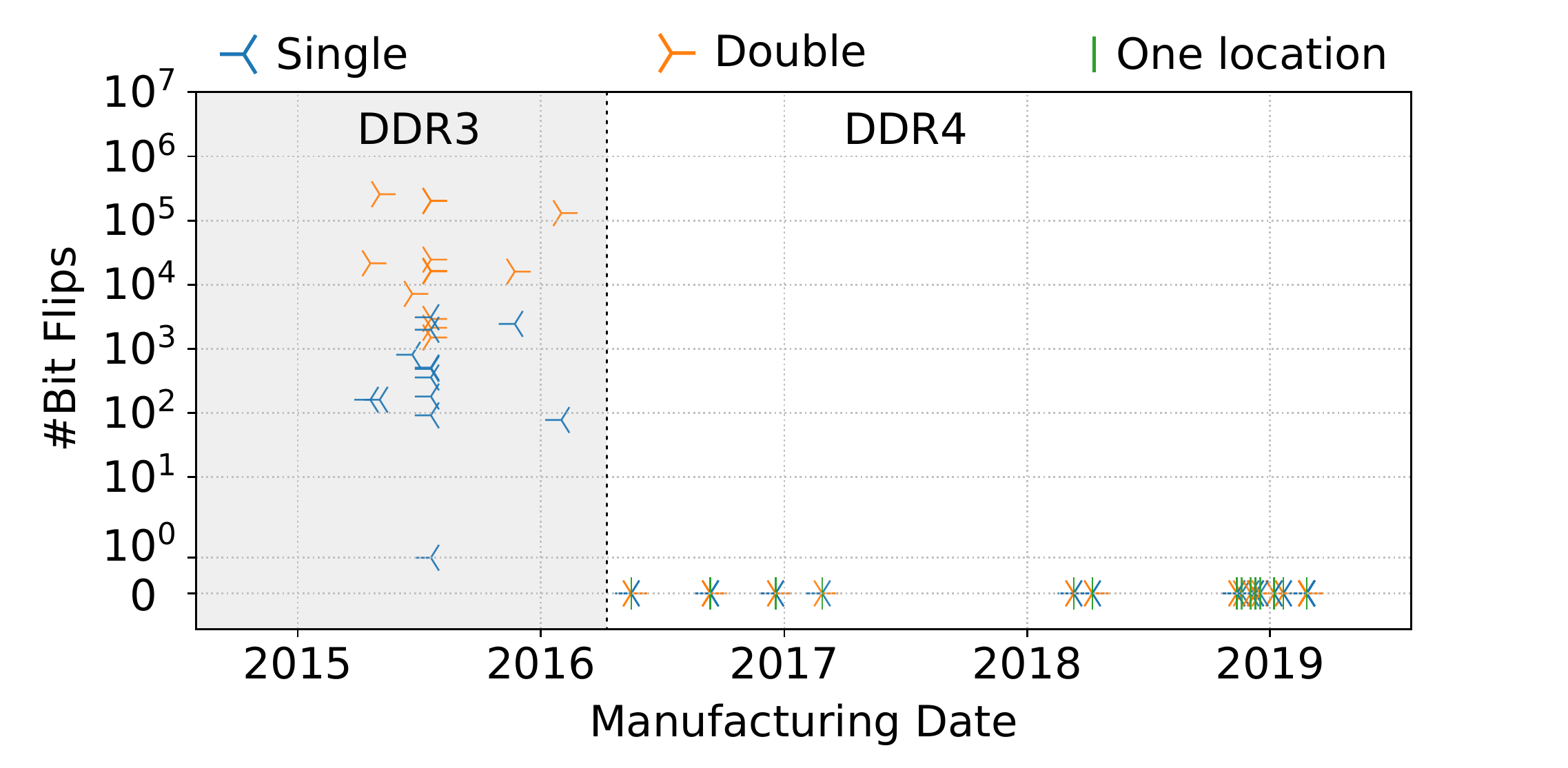}
    \caption{\textbf{Bit flips vs. manufacturing
    \hht{date}.}\protect\footnotemark{} Analysis of standard hammering
    patterns~\cite{kim2014flipping,seaborn2015exploiting-1,tatar2018defeating}
    on new DDR4 modules. We compare our results with the dataset of Tatar
    et al.~\cite{tatar2018defeating} on DDR3 modules \hht{(shown in the
    left part of the chart)}.}
	\label{fig:rh-effectiveness}
\vspace{-3mm}
\end{figure} \footnotetext{Following prior work~\cite{tatar2018defeating},
we approximate the manufacturing date with the purchase date when the
former is unavailable (Table~\ref{tbl:dimms} \hht{shows the modules for
which we applied such approximation}).}

\section{Overview}
\label{sec:overview}

We start our analysis by \hhf{showing that TRR is not a single \rh
mitigation mechanism.}
%
Specifically, we demonstrate that TRR is an umbrella term for different
solutions at different levels of the hardware stack. Next, we analyze what is
arguably the best-known TRR implementation in memory controllers, Intel
pTRR\hht{~\cite{intel2014ptrr}},
and show that it is not deployed in any consumer system \hht{we tested}
(Section~\ref{sec:rev-mc}). Since our results indicate that \hhf{recent
systems do not use TRR implemented} at the memory controller level, we \hh{analyze TRR
implementations in the DRAM chip.}
%
%
We examine in detail the \hh{effectiveness of} the TRR mitigations that different
manufacturers \hh{employ} inside their \hh{chips}. In particular, we show
that once we reach a solid understanding of the behavior of the mitigation
\hht{mechanism} and build targeted access patterns accordingly, existing
\hht{in-DRAM} \rh mitigations \hh{become} ineffective and \hh{one} can
still trigger \hh{\rh} bit flips under standard conditions (Section~\ref{sec:rev-dram}).

\hh{We observe that 1) different DRAM chips across vendors and generations can
employ different TRR implementations and 2) the distribution of DRAM cells
that are vulnerable to \rh is different for every chip.}
Since extensive investigation of every possible memory module 
\hh{is not practical}, we generalize
the insights gained from our investigation to build \n: a black-box
fuzzer for ``TRR-aware'' \rh analysis and exploitation
(Section~\ref{sec:bb-test}). We show how \n can \hht{construct} a plethora of
new and effective \rh patterns on multiple
\hht{TRR-protected} DRAM modules. \hh{We} analyze these patterns, which we collectively refer to as \manysided \rh (Section~\ref{sec:eval}), and discuss
the implications of TRR-aware hammering exploitation, showing how an
attacker armed with \n can mount successful state-of-the-art \rh attacks on
\hht{recent} DDR4 systems (Section~\ref{sec:exploitation}).


\section{Analyzing the memory controller}
\label{sec:rev-mc}

After the initial discovery of \rh~\cite{kim2014flipping,intelcorp2016row,intelcorp2016distributed,intelcorp2015method,intelcorp2014method}, BIOS vendors
\hh{first responded to the vulnerability by doubling} \hh{the} DRAM refresh
rate~\cite{refresh-lenovo, aweke2016anvil, AppleRefInc}. \hh{However, increasing the
refresh rate incurs high overhead as} more refresh \hh{operations} consume
more \hhc{energy} and \hhn{delay} actual data
transfers\hht{~\cite{kim2014flipping, liu2012raidr}}. As a
consequence, manufacturers of newer CPU generations designed and deployed
more efficient and effective hardware-based
\hh{\rh} mitigations~\cite{intelcorp2014method, intelcorp2016row,
intelcorp2016distributed, amd2016trcpage, amd2017agesa,
intel2014ptrr}---solutions that would also prevent attacks on vulnerable
DDR3 \hh{chips}.

\hh{As the \hhf{MC} services all incoming memory requests from
CPU cores, it can efficiently track the requests and implement
countermeasures in case of a \rh attack.}
Specifically, the \hhf{MC} can actively monitor the number of
activations to specific \hh{DRAM} rows and then thwart an attack by sending
additional activations to DRAM rows that might be affected by \rh. Intel's
\emph{pseudo-TRR}~\cite{intel2014ptrr} (or pTRR) is the most \hht{prominent}
example of a \rh defense that is deployed in the memory controller.
However, while it \hh{is} widely cited in the
literature~\hhn{\cite{lipp2018nethammer,gruss2018another,aweke2016anvil,seaborn2015exploiting-1,
tatar2018throwhammer, thirdio_nodate_rowhammer, vanderveen2016drammer, gruss2016rowhammer}},
very little is \hh{actually} known about \hh{the pTRR mechanism}. In this
section\hh{,} we \hht{aim to verify} the existence of \hh{pTRR} and analyze
different Intel systems to better understand the deployment \hhn{and
effectiveness} \hht{of pTRR}.


\subsection{TRR-compliant \hhn{Memory}}
\label{sec:rev-mc:spd}

To protect DRAM from \rh{} \hhf{using additional targeted refresh
operations}, the \hhf{MC} must know the \emph{maximum
number of} \activate{}s \emph{a row can bear} before \hh{any bit} in its neighboring rows
\hh{flips}. \hh{Since the discovery of \rh, manufacturers \hhf{typically} store this
information on the \emph{Serial Presence Detect} (SPD) chip\hht{~\cite{kim-isca2012}}
of the DRAM module and refer to it as \emph{Maximum Activate Count} (\mac).}
The SPD is a \hh{small read-only memory} chip containing information about
the memory module (Figure~\ref{fig:dram:high}). The CPU \hh{reads the SPD}
at boot time to gather all the necessary parameters required to
initialize the memory controller\hh{, including} the \mac field.
\hh{DRAM modules} disclosing this field have been available approximately
since 2014 and we denote them as \emph{TRR-compliant}. We discuss further
details in Appendix~\ref{apdx:trr-compliant}.

\hh{The JEDEC standard specifies} three possible configurations for \hh{the} \mac value:
\begin{enumerate*}[label=(\roman*)]
    \item \emph{unlimited}, if the \hh{DRAM module}  claims to be \rh{}-free; 
    \item \emph{untested}, if the \hh{DRAM module} was not inspected after production; or 
\item a discrete value that describes the actual number of activations the
    \hh{DRAM module} can bear (e.g., \emph{300K}).
\end{enumerate*}
We \hh{read out the \mac of} the \nDimms DDR4 modules \hh{we test}. We
\hh{find} that, regardless of the DRAM manufacturer, most \hh{of} these
\hh{modules} claim to be \rh-free by reporting an \emph{unlimited} \mac value (Table~\ref{tbl:dimms}). 



\subsection{Intel pTRR \hhn{Explained}}
\label{sec:rev-mc:ptrr}

We now take a closer look at the only publicly advertised MC-based solution
for Intel CPUs: \emph{pseudo-TRR} (or pTRR)~\cite{intel2014ptrr}.
Introduced in the Ivy Bridge EP server family~\cite{intel2014ptrr}\hh{,} pTRR refreshes victim rows
when \hh{the number of row activations issued to the DRAM exceeds the \mac
value}---according to Intel's
public documentation~\cite{intel2014ptrr}. Unfortunately, this solution is
not applicable to non-TRR-compliant \hh{modules} (i.e., those without
\hhn{a} \mac value or \mac set to \emph{untested}). As a result, when such
\hh{modules} are employed, the system defaults to double refresh mode.

\vspace{2mm}
\noindent\textbf{Observing pTRR.} We analyze the only system officially
reported to support pTRR: Xeon E5-2620 v2, with DDR3
memory~\cite{intel2014ptrr}. We disable
write-protection~\cite{jedec2014spdDDR3,jedec2015spdDDR4} on the SPD of a
DDR3 module and we perform the following two experiments.


        \mbox{}\circled{1} We overwrite the \mac value setting to two configurations\hh{:}
    \emph{untested}, simulating a non-TRR-compliant \hh{DRAM module}, and
        \emph{unlimited}. As mentioned above, when non-TRR-compliant memory
        is employed\hhn{,} the system should resort to double refresh \hht{rate,} making it
        possible to detect the mitigation via frequency analysis of the
        access latency of uncached memory reads~\cite{dram-hiccup}. Indeed, we can observe that
        with \mac value set to \emph{untested}, the system resorts to double
        refresh (Figure~\ref{fig:ptrr}).

        \definecolor{orange_plot}{RGB}{255,138,56}
\definecolor{blue_plot}{RGB}{75,145,192}
\setul{0.5ex}{0.3ex}

\begin{figure}[!hb]
    \centering
    \vspace{-2mm}
    \includegraphics[width=.9\linewidth]{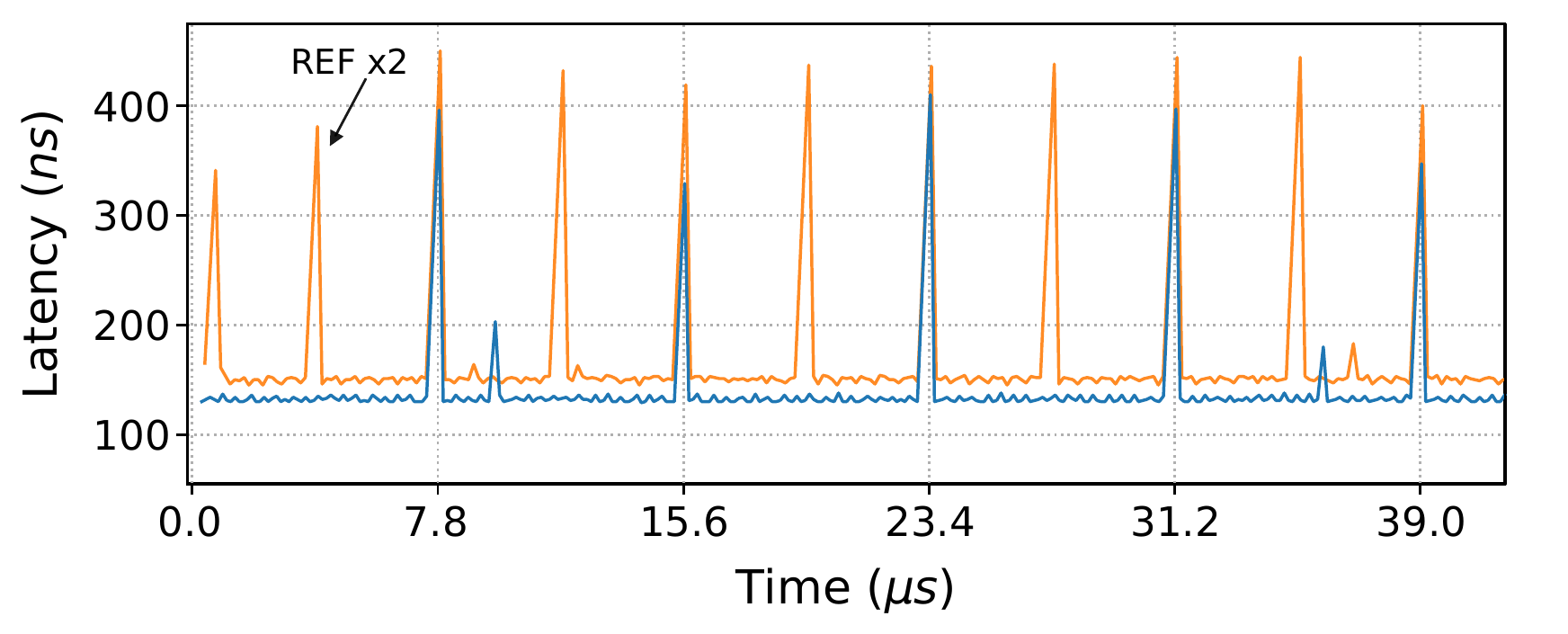}
    \vspace{-2mm}
    \caption{\textbf{Intel pTRR - \hhn{Frequency of \refresh commands}.} Uncached memory access latency with \mac value set to \ulcolored{blue_plot}{\emph{Unlimited}} and \ulcolored{orange_plot}{\emph{Untested}} on Xeon E5-2620 v2. \pp{The peaks reveal the delay introduced by the \refresh command. We observe twice as many peaks when the \mac value is set to \emph{Untested}.}}
    \label{fig:ptrr}
\end{figure}

        \mbox{}\circled{2} We overwrite the \mac value to different discrete values,
    expecting to observe a difference in the number of \hht{bit flips}. 
    \pp{In the leftmost stack of Figure~\ref{fig:ptrr-effectiveness}, we
        show the result of \hht{this} experiment when hammering the same chunk of
        memory with \mac value set to \hhn{\emph{400K} or to \emph{unlimited}}.
        \hht{We observe that} the number of bit flips drastically decreases when
        pTRR is enabled \hhn{(i.e., \mac value set to 400K)}.}
        \hh{Additionally} \hhc{(not shown in
        Figure~\ref{fig:ptrr-effectiveness})}, we discover that when setting the \mac value to
        \hhn{the minimum value defined in the DDR3
        specification~\cite{jedec2014spdDDR3}} (i.e., 200K)\hhn{,} the system
        treats the \hh{module} as a non-TRR-compliant module; that is, it
        enables double refresh. \pph{We do not analyze the effectiveness of
        pTRR in mitigating \rh \hhc{bit flips} \hhf{in this paper.} Lipp et
        al.~\cite{lipp2018nethammer} \hhc{report} bit flips on a
        \hhf{pTRR-enabled} system, and \hhf{operating at increased refresh
        rate} (i.e., \hhf{double} refresh rate) is known to be
        ineffective~\cite{aweke2016anvil,cojocar2019eccploit,kim2014flipping}
        \hhf{\hhn{at} protecting against \rh}.}




\begin{figure}[!hb]
    \centering
    \vspace{-3mm}
    \includegraphics[width=0.8\linewidth]{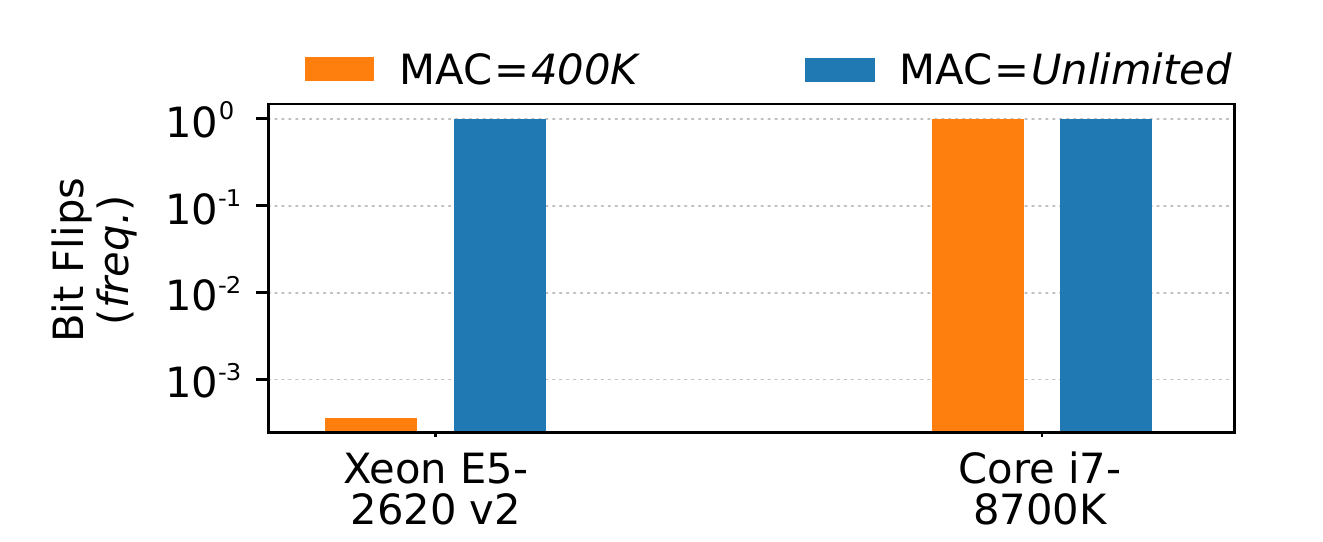}
    \vspace{-2mm}
    \caption{\textbf{Intel pTRR - Bit \hhn{flips observed with different
    \mac values}.} \pp{Frequency of observed bit flips \hhn{for} different
    \mac configurations}. Comparison between a system employing pTRR (Xeon
    E5-2620 v2) and a system with no MC-based \hh{\rh} mitigation (Core
    i7-8700K).}
    \label{fig:ptrr-effectiveness}
    \vspace{-3mm}
\end{figure}

\vspace{2mm}
\noindent\textbf{\hhn{pTRR has} limited deployment.} We run the two
experiments \hhn{for analyzing pTRR} on 6 other Intel CPUs \hh{from}
different architecture families \hh{that are descendants} of Ivy
Bridge---both server and consumer lines. Surprisingly, \hhn{the first
experiment \circled{1} shows that, when the \mac value is set to \emph{untested}, the
memory controller of each of these 6 CPUs still refreshes the DRAM with the
default (not double) refresh rate. This observation shows that
no \rh mitigation is present at the memory controller level in these CPUs.}
%
%
We corroborate this hypothesis by carrying out the second \hhn{experiment}
\circled{2} where we measure the number of bit flips \hht{as we vary} the \mac
value. \hhn{We use the new \rh patterns we present in
Section~\ref{sec:bb-test} for the CPUs that support DDR4.} 
\hhn{In contrast \hhc{to} Xeon E5-2620 v2 \hhc{server-line} CPU, which is reported to support
pTRR\hhc{~\cite{intel2014ptrr}}, the second experiment on \hhc{consumer}-line CPUs does \hhc{\emph{not}} identify a different number of bit flips
for different \mac values.}
In the rightmost stack of Figure~\ref{fig:ptrr-effectiveness}\hh{, we show}
the results for the Intel Core i7-8700K \hhc{consumer-line} \hh{CPU} as an example to
illustrate the difference between any of these consumer systems and a
pTRR-enabled system. \hhn{This experiment confirms that pTRR is in fact not
present in customer-line CPUs that we test}. We \hh{list} the deployment of
\hh{MC-based \rh} mitigations \hhc{in both server- and consumer-line CPUs} in Table~\ref{tbl:ptrr}.

\begin{table}[!h]
\centering
\caption{\textbf{Memory controller defenses.} Defenses \pp{detected in our experiments} on Intel CPUs starting from the Ivy Bridge family.}
\begin{threeparttable}
\newcommand\dgs{$^{\dagger}$}
\resizebox{\linewidth}{!}{
\begin{tabular}{ l  c c c c}
\toprule
\multicolumn{1}{c}{\emph{CPU}} & \emph{Family} & \emph{Year} & \emph{\makecell{DRAM\\generation}}  & \emph{Defense}\\
\midrule
\multicolumn{5}{l}{\textit{Server Line}}\rule[-1.2ex]{0pt}{0pt}\\
\stripe
Xeon E5-2620 v4		& Broadwell			& 2016 & DDR4	& \texttt{REF}$\times$2 \\
Xeon E5-2620 v2 	& Ivy Bridge EP		& 2013 & DDR3	& pTRR \\ 
\stripe
Xeon E3-1270 v3 	& Haswell 			& 2013 & DDR3	& \notavail  \\ 
\cmidrule(lr){1-5}
\multicolumn{5}{l}{\textit{Consumer Line}}\rule[-1.2ex]{0pt}{0pt}\\
\stripe 
Core i9-9900K 	& Coffee Lake R 	& 2018 & DDR4	& \notavail \\
Core i7-8700K 	& Coffee Lake 		& 2017 & DDR4	& \notavail \\ 
\stripe
Core i7-7700K 	& Kaby Lake 		& 2017 & DDR4	& \notavail \\
Core i7-5775C 	& Broadwell 		& 2015 & DDR3	& \notavail \\ 

\bottomrule
\end{tabular}
    } 
\end{threeparttable}
\label{tbl:ptrr}
\vspace{-2mm}
\end{table}


\subsection{Discussion}
\label{sec:rev-mc:results}

\hh{Our experiments show that the}
memory controller-based \rh mitigations are deployed only \hh{in} specific
families of \hht{Intel} processors. While we \hh{find} \hhf{that} pTRR and
other mitigations (\hhn{e.g.}, double refresh) are \hh{used in} high-end Xeon servers, our
results show that \hhf{neither DDR3 nor DDR4} consumer systems
appear to enable any MC-based mitigation. In Figure~\ref{fig:trr-timeline},
we reconstruct a timeline of \rh mitigations on Intel platforms
based on the results of our analysis. \hh{With both} DDR3 and DDR4, only
server platforms appear to benefit from mitigations inside the memory
controller while consumer platforms \hhn{do not}. \pp{\hht{Based on}
earlier reports of bit flips using standard \rh patterns on consumer DDR4
memory~\cite{gruss2018another, thirdio_nodate_rowhammer,
lipp2018nethammer}, we can speculate that \hht{in-DRAM} mitigations are
widely-deployed \hhn{only} since 2016 (i.e.,
the earliest manufacturing date \hht{of a DRAM module with \mac set to
\emph{unlimited}} among \hhn{all} modules that we list in Table~\ref{tbl:dimms}). }
In other words, DRAM manufacturers' promises of a
\rh-less future~\cite{micron2016ddr4,samsung2014trrDram} hinge entirely on
the security of their undocumented \hht{in-DRAM} TRR mitigations. Unfortunately,
as we \hh{show} in the next sections, analyzing \hhn{and understanding}
such mitigations can reveal significant weaknesses that can be exploited to
mount \rh attacks on modern DDR4 \hht{DRAM chips}.


\begin{figure}[t]        
	\centering
	\includegraphics[width=\linewidth]{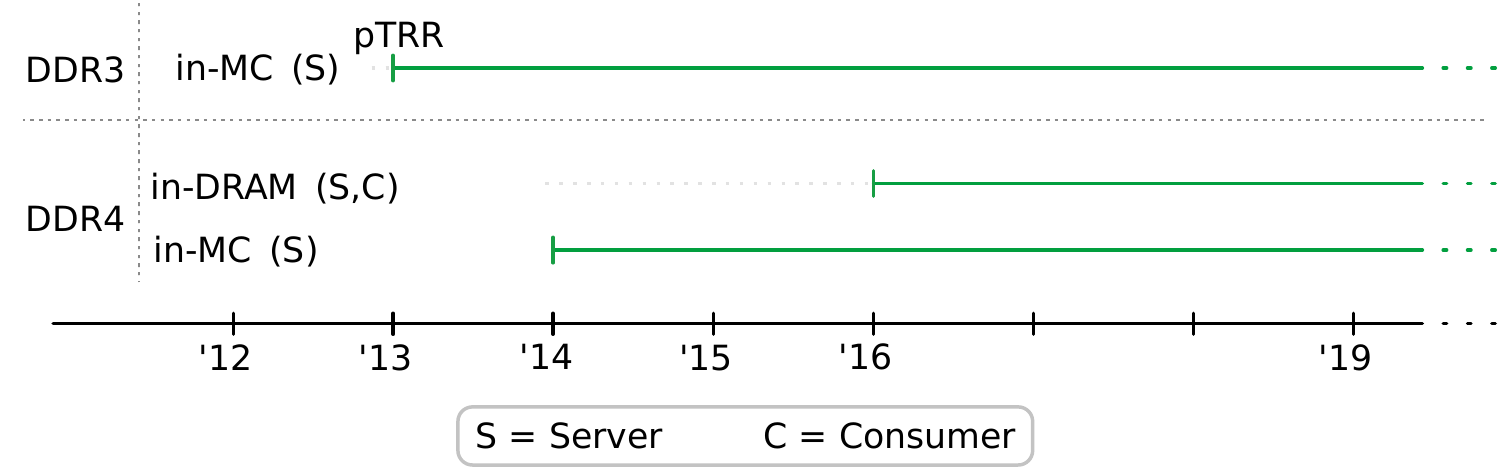}
    \caption{\textbf{TRR Timeline.} Timeline of deployment of TRR
    \hh{as \rh} mitigation. MC-based mitigations \hh{are} deployed \hhn{in} both
    DDR3 and DDR4 server systems since 2013~\cite{intel2014ptrr}. \hh{In
    contrast,} \hhn{in-DRAM} mitigations \hh{appear} with DDR4 for both consumer and server
    systems~\cite{micron2016ddr4, samsung2014trrDram}.}
    \label{fig:trr-timeline}
\vspace{-4mm}
\end{figure}

\section{Inside the DRAM chips}
\label{sec:rev-dram}

\hh{We dig deeper to understand the \rh protection that the DRAM
vendors implement inside recent DRAM chips, which \hht{are
advertised} as \rh-free~\cite{micron2016ddr4,samsung2014trrDram}}. \hh{So far, the DRAM vendors have not publicly shared the
details of the exact \rh protection \hht{mechanisms} they implement in the
form of TRR. Therefore,} 
we \hh{experiment with and analyze real DRAM chips} to shed light on the inner workings
of \hh{the TRR mechanisms implemented by different vendors \hht{in} different DRAM chip
generations}. \hh{Performing such an analysis using a general-purpose CPU
is extremely challenging because the memory controller provides a very high-level 
interface to the CPU (i.e., the
programmer can interface \hht{with} the DRAM using only
load/store instructions). However, to perform accurate 
experiments, we need fine-grained control over the low-level commands sent
to the DRAM. Therefore,} 
we leverage \hh{an open-source FPGA-based memory controller,
SoftMC~\hhn{\cite{hassan2017softmc, softmc-github}}, which enables the programmer to issue
arbitrary DRAM commands in a cycle-accurate manner. \hhn{We} extend SoftMC to
support experimental studies on DDR4 modules.} \hh{We first discuss our hypotheses for potential ways of implementing
in-DRAM TRR. Then,} we present case studies for two DRAM modules from
different manufacturers. Our results show that different manufacturers implement vastly different TRR mitigations. 

\subsection{Building Blocks and Hypotheses}
\label{sec:rev-dram:preliminaries}

%
%
While literature indicates that each manufacturer may implement its own
variant \hht{of TRR}~\hhn{\cite{microntechnologyinc2018apparatuses,
skhynixinc2017refresh, skhynixinc2017memory, skhynixinc2016smart,
micron2017rowCopy, samsung2016preciseRowTRR, skhynix2016trrStorageUnit,
samsung2017mcTRR}}, we abstract the implementation details and unravel the
two main requirements for supporting TRR\hhn{: the sampler and the
inhibitor}. We define these requirements as building blocks and present a
series of hypotheses that we verify in the next sections.

\vspace{3mm}
\noindent\textbf{The \hhn{Sampler}.} A sampling mechanism is required to track
\hht{which} aggressor rows are being hammered. Solutions vary from basic
frequency-based sampling to more \hht{complex} designs that track
activations per row. In frequency-based implementations, sampling occurs at
fixed periods in time within a refresh
interval~\cite{microntechnologyinc2018apparatuses,
skhynix2016trrStorageUnit, skhynixinc2016smart}.  For example, a TRR
implementation may determine aggressor rows by monitoring every
3\textsuperscript{rd} and 4\textsuperscript{th} access after a \refresh.
The more complex designs that track accesses on a per-row basis, keep
activation counters for \hht{a number of}
rows~\cite{samsung2017mcTRR,samsung2016preciseRowTRR} and select aggressors
based on their individual activation counts. Despite differences in
\hht{its} implementation, the goal of the sampler remains the same: track which
rows are being hammered in order to identify their \textit{target} victim
rows.\\

\vspace{-2mm}
Our first hypothesis is that the sampler has a limited size $s$. In other
words, there is a maximum number of aggressor rows \hhn{that the sampler} can track. Phrased
differently, the \hhn{TRR} mitigation can protect only a limited number of victim
rows.

\vspace{3mm}
\noindent\textbf{The Inhibitor.} Once the sampler is aware of the aggressor
rows, the mitigation must thwart the hammering process. As the name
\emph{Target Row Refresh} suggests (and different designs
confirm~\cite{microntechnologyinc2018apparatuses, skhynixinc2016smart,
skhynix2016trrStorageUnit}), an effective solution consists of generating
extra refreshes for the victim rows. Nonetheless, more \hht{sophisticated} designs
\hhn{incorporate} the possibility of row
remapping~\hhc{\cite{micron2017rowCopy, liu2012raidr, khan2016parbor}}.\\

\vspace{-2mm}
Our second hypothesis is that the inhibitor acts at refresh time---based on the literature~\cite{microntechnologyinc2018apparatuses, skhynixinc2016smart, skhynix2016trrStorageUnit}. 
\hhn{Refresh} operation is the
responsibility of the memory controller\hhn{,} which issues \hhn{one \refresh
command} every $7.8\,\mu{}s$ (\trefi).
Since DDR is a synchronous protocol~\cite{jedec2014ddr4,jedec2014lpddr4},
the memory controller must remain idle for a fixed period of time (\trfc)
before it can send subsequent commands to the bank. Any, possibly
additional, targeted refreshes must still respect these timing
constraints for the \hhn{DRAM module} to be compliant \hhc{with} \hhn{the DDRx
standard}. That is, only a limited number of \hht{target} rows can be
refreshed.

\vspace{3mm}
\noindent\textbf{Goals.} Based on the aforementioned assumptions we define the following questions that we want to answer.
\begin{itemize}
\item \hhn{What is the size of} the sampler?
\item How does the sampler track aggressor rows? For example, does it record row activation commands at a constant frequency or based on a function of time? 
\item How does the inhibitor work? Can it prevent \hht{bit flips}?
\end{itemize}

In the following, we try to answer these questions by analyzing TRR
\hht{via} two different case studies.

\subsection{Case I: \hht{Module} \hnx{12}}
\label{sec:rev-dram:hnx}


Our first study \hhn{examines} a \hht{module} from manufacturer \hnx{}.  We first find
the minimum number of activations that are required to trigger \hht{bit
flips} on this module. \hhn{To do so}, we disable refresh\hhn{,} \hhf{which prevents TRR
from performing refresh on victim rows,} and perform a
double-sided \rh sweep of a single DRAM bank. The results, plotted in
Figure~\ref{fig:act-vs-flips}, show that we can trigger \hht{bit flips} with
\hht{as few as} 50K activations. This indicates that \hhf{DRAM} cells
\hhf{in DDR4} are \hhf{generally} considerably weaker
compared to \hhf{DRAM cells in the predecessor} DDR3 \hhf{standard}\hhn{,
which requires} at least \tildesmall{}139K
activations~\cite{kim2014flipping}. Nonetheless, for future experiments
\hht{reported in this paper} we
use a higher activation count so that we can observe more
\hht{bit flips} and draw stronger conclusions. 

\begin{figure}[!ht]
    \centering
    \includegraphics[width=\linewidth]{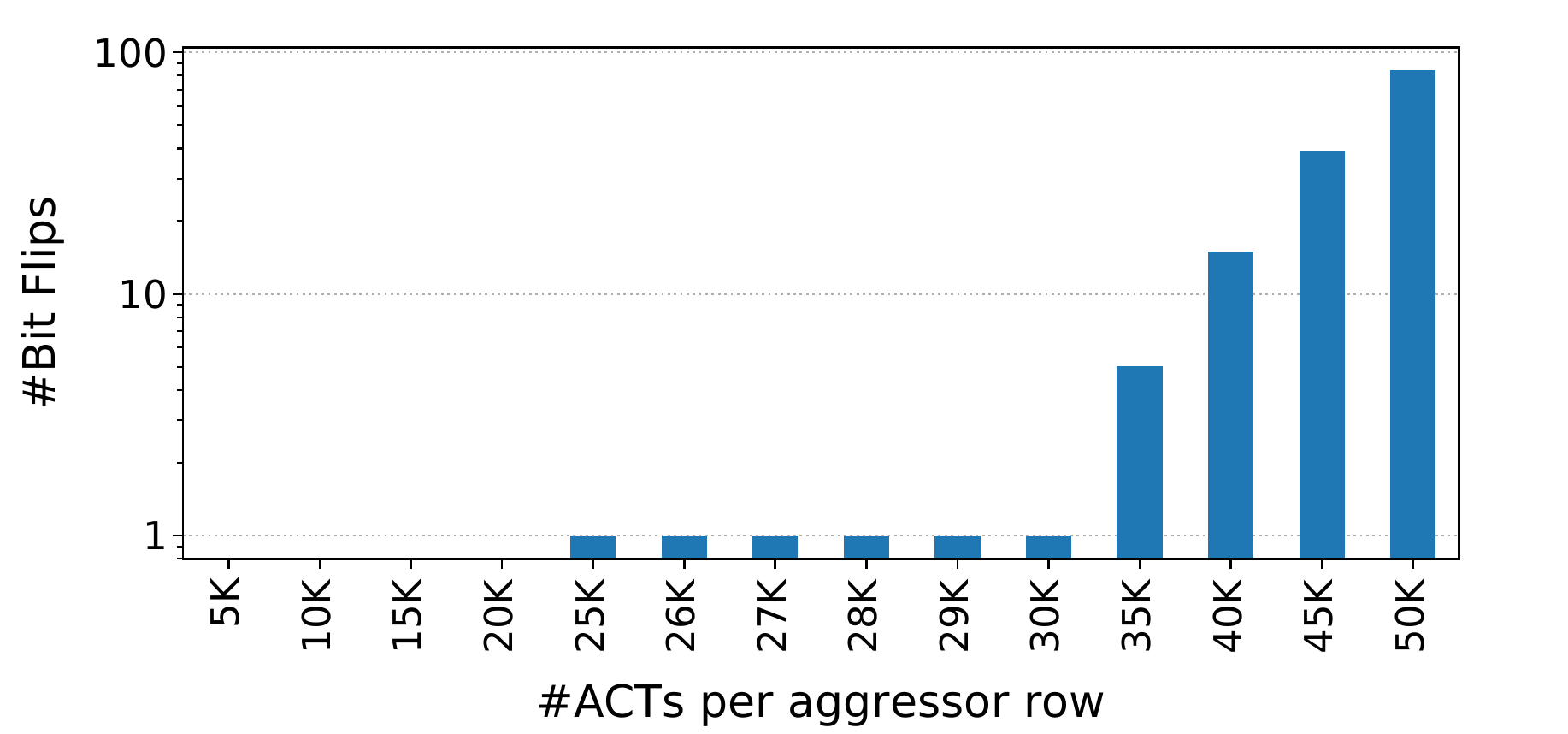}
    \caption{\textbf{\hht{Bit flips} vs. \hhn{number of activations}.} \hht{Module} \hnx{12}: We
    can observe \hht{bit flips} with as \hht{few} as 25K activations per
    aggressor row (i.e., 50K activations in total \hht{due to double-sided
    hammering}).} 
    \vspace{-4mm}
    \label{fig:act-vs-flips}

\end{figure}

\vspace{3mm}
\noindent\textbf{Mastering refresh.} Knowing the physical limitations
of the \hhn{DRAM module}, we now reintroduce the \refresh command. We decide to batch
refresh operations together with the goal of understanding the relationship
between them and the \hhn{effectiveness of the TRR} mitigation. We perform a series of hammers (i.e.,
activations of aggressors) followed by $r$ refreshes for ten rounds---we carry out 8K hammers per round. In
Figure~\ref{fig:hnx:refs-vs-flips}, we report the results of this experiment
for \rh configurations with different numbers of aggressors. Let us first
consider only the third column of the plot: double-sided \rh. We observe
that adding a single refresh causes the number of \hht{bit flips} to drop
from 2,866 to only one \hhn{(for $r = 1$), and then to zero (for any $r\geq 2$)}.
This experiment provides an insightful result: since sending multiple
\refresh commands varies the number of \hht{bit flips}, the \hht{TRR}
mitigation must act on every refresh command.\\

\vspace{-2mm}
\begin{tcolorbox}[top=0pt,left=0pt,bottom=0pt,right=0pt,boxsep=5pt,arc=0pt]
    \observation{1} The \hht{TRR} mitigation acts (i.e., carries out a targeted refresh) on \textbf{every} refresh command.
\end{tcolorbox}

\begin{figure}[b]
    \centering
    \vspace{-4mm}
    \includegraphics[width=\linewidth]{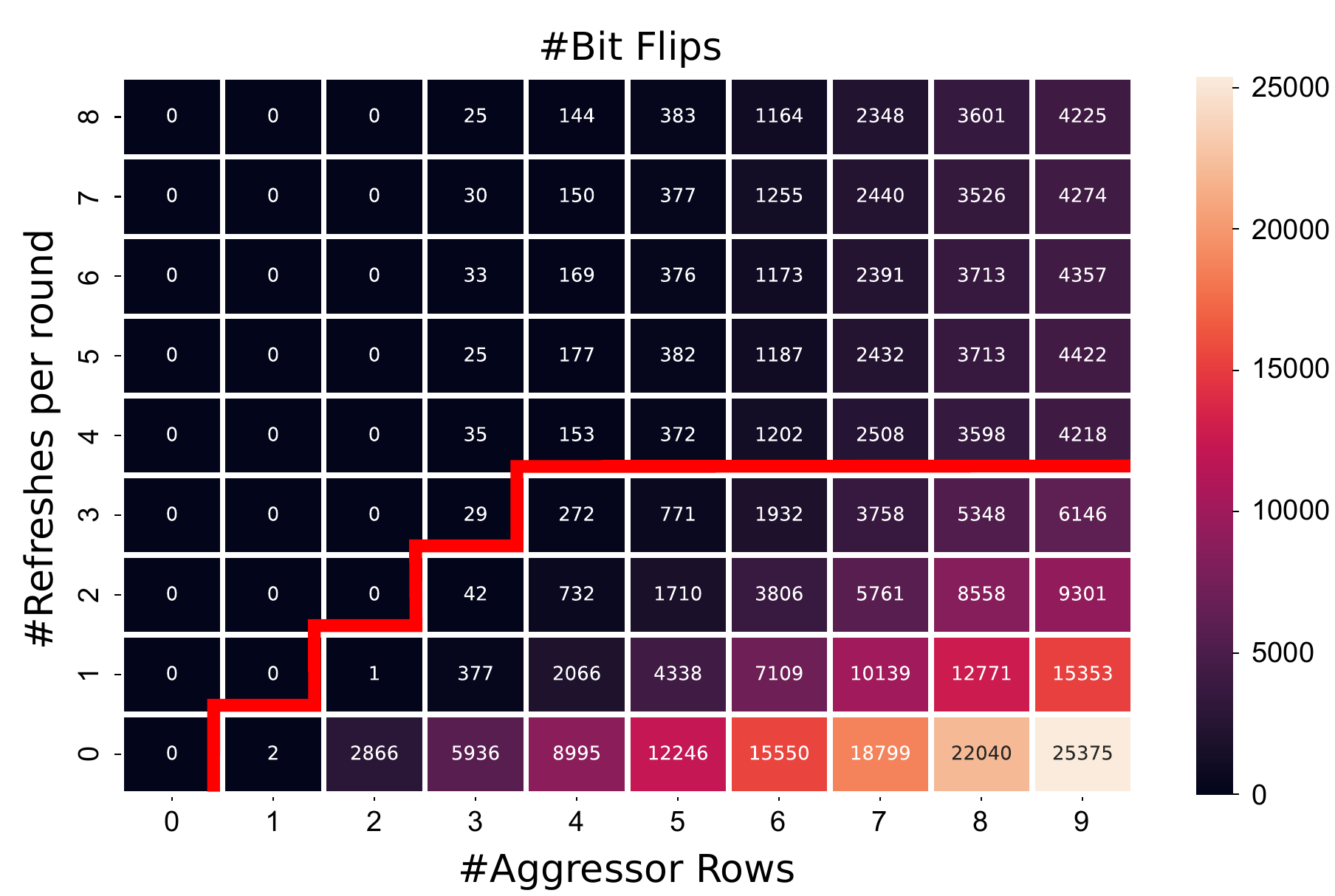}
    \caption{\textbf{Refreshes vs. \hht{Bit Flips}.} \hht{Module} \hnx{12}:
    Number of \hht{bit flips} detected when sending $r$ refresh commands to the module. We report this for different number of aggressor rows ($n$). For example, when hammering 5 rows, followed by sending 2 refreshes, we find 1,710 bit flips.   
    This figure shows that the number of \hht{bit flips} stabilizes for
    $r\geq4$, \hht{implying that} the size of the sampler \hht{may} be 4.}
    \label{fig:hnx:refs-vs-flips}
    \vspace{-3mm}
\end{figure}

Next, we take a closer look at the sampler size $s$ to find how many rows
the mitigation can handle. We increase the number of aggressors $n$ while
keeping the number of {\act}s per \hht{aggressor} row constant. For every additional aggressor
row, we have an additional victim row. For example, with 3 aggressor rows,
the hammering configuration looks like \verb+VAVAVAV+ where \verb+V+ stands
for a victim row and \verb+A+ stands for an aggressor row. The fourth
column in Figure~\ref{fig:hnx:refs-vs-flips} shows the behavior when
hammering three aggressors ($n=3$). Here we observe something different:
the number of \hht{bit flips} decreases significantly when introducing up
to two refreshes. However, it plateaus \hhc{for} $r\geq 3$ \emph{without}
\hhn{going down to zero}. Notice that when hammering both 2 and 3
rows\hhn{,} the plateaus \hhn{happen} when $r=n$. This suggests that
\emph{the \hhn{TRR} mitigation
samples more than one aggressor within a refresh interval while it can
refresh only one victim per refresh operation}. \hhf{The DRAM can refresh
only one of the victim rows} likely \hhf{as} a
consequence of the tight timing constraints imposed by the \trfc parameter.
\hhn{Moreover}, we can deduce \hht{from} the remaining \hhn{non-zero} \hht{bit flips} that the sampler
is likely to discard the aggressor row from its table once one of its
victims has been refreshed. We can recover the size of the sampler by
performing the same experiment for different numbers of aggressors $n$.
While increasing $n$, we search for the scenario where the number of
\hht{bit flips} stabilizes for $r<n$. When this happens, we can conclude
that we have \hhc{overflowed} the sampler.  We show the results of this
experiment for different values of $n$ in
Figure~\ref{fig:hnx:refs-vs-flips}. As speculated, we see the number of
\hht{bit flips} leveling \hht{off} (i.e., \hht{remaining} constant on the
y-axis) for $r\geq4$, revealing the size of the sampler \hhn{to be $s = 4$: the sampler in this module can track only 4 aggressor rows}.\\

\begin{tcolorbox}[top=0pt,left=0pt,bottom=0pt,right=0pt,boxsep=5pt,arc=0pt]
  	\observation{2} The mitigation can sample \textbf{more than one} aggressor per refresh interval.\\
  	\observation{3} The mitigation can refresh only a \textbf{single}
    victim within a \hht{refresh operation (i.e., time \trfc)}.\\
    \hhn{\observation{4} Sweeping the number of refresh operations and aggressor
    rows \hhc{while hammering} reveals the sampler size.}
\end{tcolorbox}

Based on these observations, we conclude that hammering more than 4 rows
should circumvent the mitigation. We confirm this by running a test on our
\hhn{SoftMC} FPGA infrastructure\hhc{~\cite{hassan2017softmc}} with standard conditions (i.e., \trefi $=7.8\,\mu{}s$).
Indeed, Figure~\ref{fig:hnx:std-refi} shows that we overwhelm the
mitigation \hht{by} hammering 5 rows. Figure~\ref{fig:hnx:std-refi}
provides another insight: it shows that for every \hht{number} of aggressors $>$5,
the number of \hht{bit flips} decreases drastically compared to \nsided{5}
\rh{}---suggesting that the sampler selects rows in a specific fashion.
While we tried to understand this behavior of the sampler, the lack of
visibility inside the DRAM chip made it challenging. Regardless, this
additional information is not necessary given that hammering 5 aggressors
in standard conditions already bypasses the in-DRAM mitigation. 

\begin{figure}[!ht]
    \centering
    \includegraphics[width=.9\linewidth]{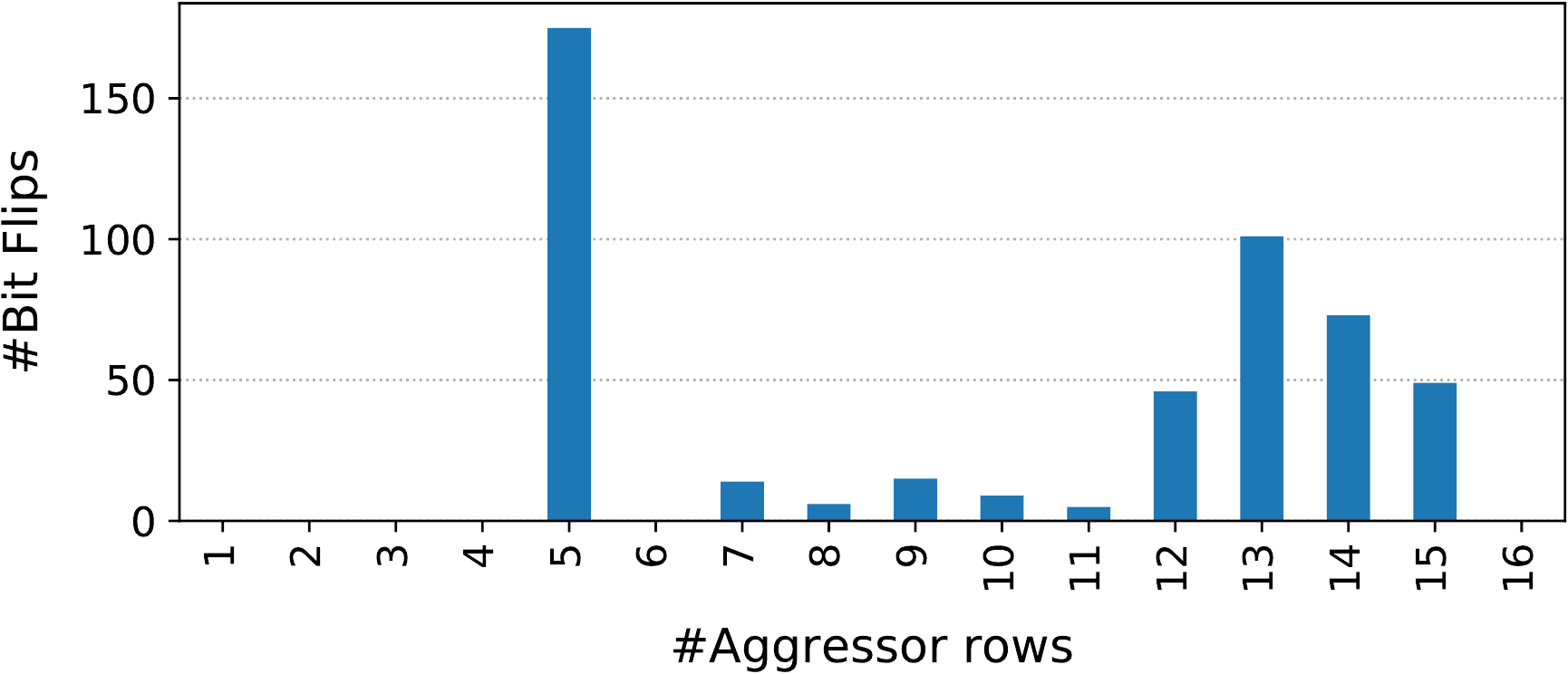}
    \caption{\textbf{\hhn{Bit flips} vs. \hhn{number of aggressor rows.}} \pp{\hht{Module}
    \hnx{12}: Number of bit flips in bank 0 \hht{as we vary the} number of
    aggressor rows. \hhf{Using SoftMC, we refresh DRAM with standard
    \trefi and run the tests until each aggressor rows is hammered 500K
    times.}}}
    \label{fig:hnx:std-refi}
\end{figure}

\subsection{Case II: \hht{Module} \smg{15}}
\label{sec:rev-dram:samsung}

To \hhn{provide an} understanding of the different flavors of in-DRAM TRR, we
further study the behavior of a memory module from a different
manufacturer: \smg{15}. We quickly test and confirm that the mitigation
acts at every refresh command, corroborating the observation made in the
previous case study. We then move to analyzing the relationship between
the number of \hht{bit flips} and the number of aggressors $n$\hhn{,} \hht{with
the default refresh rate}, depicted in Figure~\ref{fig:smg:std-refi}. We find that we can
reliably flip bits for $n\geq7$, indicating a sampler of size 6. 

\begin{figure}[t]
    \centering
    \vspace{-4mm}
    \includegraphics[width=.9\linewidth]{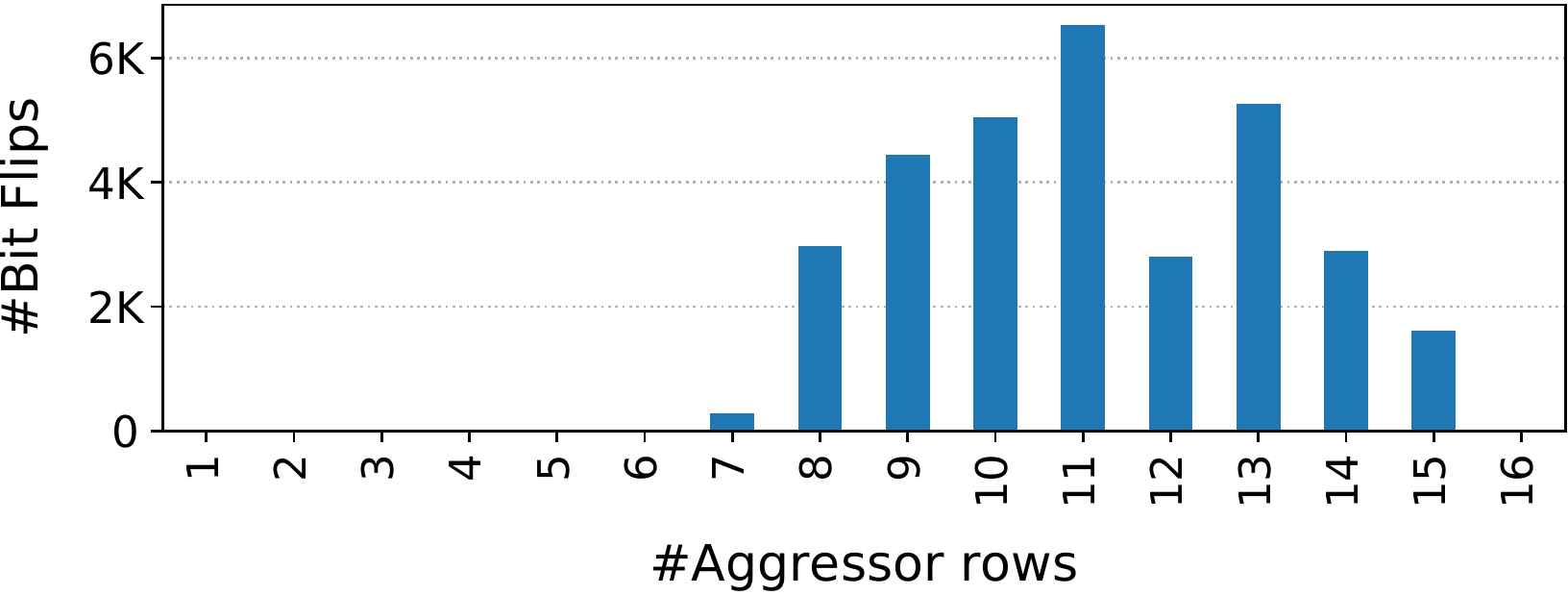}
    \caption{\textbf{Bit flips vs. \hhn{number of aggressor rows}.} Module \smg{15}: \pp{Number
    of bit flips in bank 0 as we vary the number of aggressor rows.
    \hhf{Using SoftMC, we refresh DRAM with standard \trefi and run the
    tests until each aggressor rows is hammered 500K times.}}}
    \label{fig:smg:std-refi}
	\vspace{-5mm}
\end{figure}

\noindent\textbf{Double-sided \hhc{\rh} resurrected.}
Although we now already bypass the mitigation, we take this one step
further and try to analyze the sampler to see if we can revive the more
efficient double-sided \rh attack. Our approach consists of finding the
minimal set of \textit{dummy} rows that allows us to trick the mitigation
\hht{mechanism} into refreshing \hht{all other neighbors of the hammered
rows} but our victim. For this, we \hht{focus} on a single row \hht{that} we know to be
susceptible to \hht{bit flips} and \hht{for} which we \hhf{find in advance} the threshold of
hammers required to observe bit flips. Based on this threshold, we carry
out successive experiments while modifying two parameters: (i) the
distribution of activations \hhn{across} aggressor and dummy \hht{rows} and (ii) the
number of dummies starting from 6 (i.e., the supposed size of the sampler).
To our surprise, regardless of the configuration, we could not detect any
bit flip. 

Investigating further, we discover two more parameters that were previously unaccounted for:

\vspace{1mm}
\setlength{\leftskip}{3mm}
\noindent\emph{\hhf{DRAM command order} dependency.} The sampler may act at
specific \hhf{DRAM commands issued}
within a refresh interval and \hhn{thus it may} not necessarily
\hhn{sample} based \hhn{only}
on frequency. In the case of \hht{module} \smg{15}, \hht{the sampler} seems to record the first $\alpha$
activations after a refresh command---where \hhn{$\alpha \leq 6$}.

\vspace{1mm}
\noindent\emph{Address dependency.} In \hht{module} \smg{15}, we observe a dependency
between the aggressor-row address and the \hht{dummy rows'} addresses. That is, when
hammering two aggressor rows, we detect more \hht{bit flips} when
\hht{we pick particular} dummy rows \hht{compared to picking random dummy
rows}. This suggests \hhn{that the} design of the sampler \hhn{involves} 
\hhf{optimizations \hhc{to reduce} \hhn{the} storage cost \hhn{of tracking row
activations} such that}
multiple aggressor \hht{rows' addresses} may conflict. \\

\setlength{\leftskip}{0mm}
\begin{tcolorbox}[top=0pt,left=0pt,bottom=0pt,right=0pt,boxsep=5pt,arc=0pt]
    \observation{5} \hhc{The sampler records row activations at specific
    commands and likely at specific ordering of commands} (i.e.,
    \hht{it performs} \textbf{\hhf{command-order}-based sampling}).\\
    \observation{6} The \hht{sampling mechanism is affected by the
    addresses of aggressor rows (i.e., \hhn{sampling is} \textbf{row-address-dependent})}. 
\end{tcolorbox}

\subsection{Running on the CPU: \hht{Module} \smg{15}}
\label{sec:rev-dram:cpu-samsung}

While we observe a considerable \hht{number} of bit flips when \hht{we use} the
(optimal) activation pattern \hht{discovered by} SoftMC, a custom FPGA memory
controller does not represent a widespread \hhn{threat} model. As a
consequence, we want to \hht{check if we can} reproduce the same access
pattern when running on commodity hardware, such as a regular desktop
computer. 

During the analysis process, we \hht{find} the mitigation of the \smg{15}
memory module to be \hhf{command order} and address dependent. This represents a great
challenge when trying to reproduce access \hhn{patterns that cause bit
flips} from the CPU. In
fact, in order to fool the mitigation, we need to carry out a specific
series of activations right after a \refresh command to keep the inhibitor
busy with another set of rows \hhn{than the intended victim row}. This means we need to synchronize our access
pattern with the \refresh command. \hht{Even though we can} detect refresh
operations (Section~\ref{sec:rev-mc}), synchronizing our access pattern
with them is much more \hht{difficult}. 
We \hhf{re-implement} the access \hhf{patterns} discovered in the analysis
process\hhn{, which we explain in Section~\ref{sec:rev-dram:samsung},} to
run on the CPU. However, we \hhf{observe} \hht{much fewer} bit flips
compared to what we \hhf{obtain} with SoftMC, suggesting we may not be able to
perfectly synchronize \hht{the hammering} pattern \hht{with the refresh
operations using a CPU}. This is likely due to the fact that the memory controller
applies various optimizations that can reorder memory requests and refresh commands.

\subsection{Observations}
\label{sec:rev-dram:discussion}

Our experiments \hhn{in Section~\ref{sec:rev-dram:cpu-samsung}} show the
difficulty of reproducing our FPGA results---those obtained in a
simplified, controlled environment---on a modern CPU. This advocates for a
better solution for finding effective access patterns that trigger \hht{bit
flips} on TRR-protected DDR4 chips. In the next section, we introduce \n,
a black-box \rh test suite that generates effective access patterns to
bypass in-DRAM TRR solutions.      

\n is inspired by the insights that we obtained using our analysis of TRR-protected DDR4 chips in this section. More specifically, we take advantage of the following
\hht{insights}:\\ 

\vspace{-2mm}
\begin{enumerate}[label=${\arabic*}$.]
    \item The sampler can track a limited number of \hht{aggressor rows}. Thus, we
        may need to \textit{\hhn{overflow}} the sampler's aggressor \hht{rows} \emph{table} in order
        to bypass the \hht{TRR} mitigation.
    \item The sampler may sample activations at specific
        \hhf{commands}\hhn{, at a specific} frequency\hhn{, or both}.
    \item The sampler \hhn{design} may be \hhn{row} address dependent.
        Therefore, some rows may be easier to hammer than others \hhn{and}
        the same \hhn{set of} rows activated in different order may yield completely different results. 
    \item The cells \hht{in DDR4 chips} are much \hhn{more \rh-prone} than
        \hhn{those} on DDR3~\cite{kim2014flipping}, \pp{requiring
        \hht{fewer} activations to trigger bit flips.}
\end{enumerate}

\par In the next section, we describe how we use these observations to build a
(guided) black-box fuzzer \hht{that can cause} bit flips on TRR-protected DDR4 modules.

\begin{table*}[!t]
\newcommand\dgs{$^{\dagger}$}
\newcommand\ddgs{$^{\ddagger}$}
\newcommand\dia{$^{\diamond}$}
\newcommand\nt{\texttt{N/A}}

\centering
\caption{\textbf{\n results.} We report the number of patterns found and bit flips detected for the \nDimms DRAM modules in our set.}

\begin{threeparttable}

\resizebox{\linewidth}{!}{
\begin{tabular}{ l l c c c c c c c r c c c c}
\toprule
\mr{2.5}{\emph{Module}} & \mr{2.5}{\emph{\makecell{Date\\(yy-ww)}}} & \mr{2.5}{\emph{\makecell{Freq.\\(MHz)}}} & \mr{2.5}{\emph{\makecell{Size\\(GB)}}} & \multicolumn{3}{c}{\emph{Organization}} &  \mr{2.5}{\emph{MAC}} & \mr{2.5}{\emph{\makecell{Found\\ Patterns}}} &  \mr{2.5}{\emph{\makecell{Best Pattern}}} & \multicolumn{3}{c}{\emph{Corruptions}} & \mr{2.5}{\emph{\makecell{Double\\Refresh}}}\\
\cmidrule(lr){5-7}\cmidrule(lr){11-13}
\multicolumn{4}{c}{} 				   & \emph{Ranks} & \emph{Banks} & \emph{Pins} & & & &  \emph{Total} & $1\to0$ & $0\to1$  \\
\midrule

\stripe
\smg{0,1,2,3}  & \ 16-37          & 2132  & 4         & 1     & 16    & $\times$8   & \texttt{UL}  & \notavail  & \notavail         & \notavail    & \notavail    & \notavail  & \notavail  \\
\smg{4}        & \ 16-51          & 2132  & 4         & 1     & 16    & $\times$8   & \texttt{UL}  & 4          & \nsided{9}        & 7956         & 4008         & 3948       & \notavail  \\
\stripe
\smg{5}        & \ 18-51          & 2400  & 4         & 1     & 8     & $\times$16  & \texttt{UL}  & \notavail  & \notavail         & \notavail    & \notavail    & \notavail  & \notavail   \\
\smg{6,7}      & \ 18-15          & 2666  & 4         & 1     & 8     & $\times$16  & \texttt{UL}  & \notavail  & \notavail         & \notavail    & \notavail    & \notavail  & \notavail  \\
\stripe
\smg{8}        & \ 17-09          & 2400  & 8         & 1     & 16    & $\times$8   & \texttt{UL}  & 33         & \nsided{19}       & 20808        & 10289        & 10519      & \notavail   \\
\smg{9}        & \ 17-31          & 2400  & 8         & 1     & 16    & $\times$8   & \texttt{UL}  & 33         & \nsided{19}       & 24854        & 12580        & 12274      & \notavail  \\
\stripe
\smg{10}       & \ 19-02          & 2400  & 16        & 2     & 16    & $\times$8   & \texttt{UL}  & 488        & \nsided{10}       & 11342        & 1809         & 11533      & \checkmark  \\ 
\smg{11}       & \ 19-02          & 2400  & 16        & 2     & 16    & $\times$8   & \texttt{UL}  & 523        & \nsided{10}       & 12830        & 1682         & 11148      & \checkmark  \\ 
\stripe
\smg{12,13}    & \ 18-50          & 2666  & 8         & 1     & 16    & $\times$8   & \texttt{UL}  & \notavail  & \notavail         & \notavail    & \notavail    & \notavail  & \notavail  \\
\smg{14}       & \ 19-08\dgs      & 3200  & 16        & 2     & 16    & $\times$8   & \texttt{UL}  & 120        & \nsided{14}       & 32723        & 16490        & 16233      & \notavail  \\
\stripe
\smg{15}\ddgs  & \ 17-08          & 2132  & 4         & 1     & 16    & $\times$8   & \texttt{UL}  & 2          & \nsided{9}        & 22397       & 12351        & 10046      & \notavail \\
\cmidrule(lr){1-14}
\mcr{0}        & \ 18-11          & 2666  & 16        & 2     & 16    & $\times$8   & \texttt{UL}  & 2          & \nsided{3}        & 17           & 10           & 7         & \notavail  \\
\stripe
\mcr{1}        & \ 18-11          & 2666  & 16        & 2     & 16    & $\times$8   & \texttt{UL}  & 2          & \nsided{3}        & 22           & 16           & 6          & \notavail  \\
\mcr{2}        & \ 18-49          & 3000  & 16        & 2     & 16    & $\times$8   & \texttt{UL}  & 2          & \nsided{3}        & 5            & 2            & 3          & \notavail  \\
\stripe
\mcr{3}		   & \ 19-08\dgs      & 3000  & 8         & 1     & 16    & $\times$8   & \texttt{UL} & \notavail  & \notavail         & \notavail    & \notavail    & \notavail & \notavail  \\
\mcr{4,5}      & \ 19-08\dgs      & 2666  & 8         & 2     & 16    & $\times$8   & \texttt{UL}  & \notavail  & \notavail         & \notavail    & \notavail    & \notavail  & \notavail  \\
\stripe
\mcr{6,7}      & \ 19-08\dgs      & 2400  & 4         & 1     & 16    & $\times$8   & \texttt{UL}  & \notavail  & \notavail         & \notavail    & \notavail    & \notavail   & \notavail  \\
\mcr{8}\dia    & \ 19-08\dgs      & 2400  & 8         & 1     & 16    & $\times$8   & \texttt{UL}  & \notavail  & \notavail         & \notavail    & \notavail    & \notavail  & \notavail  \\
\stripe
\mcr{9}\dia    & \ 19-08\dgs      & 2400  & 8         & 1     & 16    & $\times$8   & \texttt{UL}  & 2          & \nsided{3}        & 12            & \notavail   & 12         & \checkmark  \\
\mcr{10,11}    & \ 16-13\dgs      & 2132  & 8         & 2     & 16    & $\times$8   & \texttt{UL}  & \notavail  & \notavail         & \notavail    & \notavail    & \notavail   & \notavail  \\
\cmidrule(lr){1-14}
\stripe
\hnx{0,1}      & \ 18-46          & 2666  & 16        & 2     & 16    & $\times$8   & \texttt{UL}  & \notavail  & \notavail         & \notavail    & \notavail    & \notavail   & \notavail  \\
\hnx{2,3}      & \ 19-08\dgs      & 2800  & 4         & 1     & 16    & $\times$8   & \texttt{UL}  & \notavail  & \notavail         & \notavail    & \notavail    & \notavail   & \notavail  \\
\stripe
\hnx{4,5}      & \ 19-08\dgs      & 3000  & 8         & 1     & 16    & $\times$8   & \texttt{UL}  & \notavail  & \notavail         & \notavail    & \notavail    & \notavail  & \notavail   \\
\hnx{6,7}      & \ 19-08\dgs      & 3000  & 16        & 2     & 16    & $\times$8   & \texttt{UL}  & \notavail  & \notavail         & \notavail    & \notavail    & \notavail   & \notavail  \\
\stripe
\hnx{8}        & \ 19-08\dgs      & 3200  & 16        & 2     & 16    & $\times$8   & \texttt{UL}  & \notavail  & \notavail         & \notavail    & \notavail    & \notavail   & \notavail  \\
\hnx{9}        & \ 18-47          & 2666  & 16        & 2     & 16    & $\times$8   & \texttt{UL}  & \notavail  & \notavail         & \notavail    & \notavail    & \notavail   & \notavail  \\
\stripe
\hnx{10,11}    & \ 19-04          & 2933  & 8         & 1     & 16    & $\times$8   & \texttt{UL}  & \notavail  & \notavail         & \notavail    & \notavail    & \notavail   & \notavail \\
\hnx{12}\ddgs  & \ 15-01\dgs      & 2132  & 4         & 1     & 16    & $\times$8   & \texttt{UT} & 25  	  & \nsided{10}       & 190037    		& 63904    	   & 126133   &  \checkmark  \\
\stripe
\hnx{13}\ddgs  & \ 18-49      	& 2132  & 4         & 1     & 16    & $\times$8   & \texttt{UT} & 3  	  & \nsided{9}        & 694    			& 239   	   & 455   	  &  \notavail  \\

\bottomrule

\end{tabular}
} 

    \vspace{1mm}

    \scriptsize
$\dagger$\quad The module does not report manufacturing date. Therefore, we report purchase date as an approximation. \hfill \texttt{UL} = Unlimited \\
$\ddagger$\quad Analyzed \hhn{using} the FPGA-based \hhn{SoftMC}. \hfill \texttt{UT} = Untested \\
$\diamond$\quad The system runs with double refresh frequency in standard
    conditions. We configured the refresh interval to be $64\,ms$ in the
    BIOS settings.\hfill \mbox{}
\end{threeparttable}
\label{tbl:dimms}
\end{table*}


\section{\n: A TRR-aware Rowfuzzer}
\label{sec:bb-test}

To convert the knowledge that we gathered from \hhn{our} analysis process
\hhn{on \hhc{the} FPGA-based SoftMC \hhc{platform}} into
practical attacks that we can launch from regular software on \hhc{a} CPU, we
developed a guided black-box fuzzer for \rh called \n. When searching for
usable access patterns, a \hhn{CPU-based} fuzzer has two main advantages over an FPGA-based
approach: (i)~it allows an attacker to completely ignore the memory
controller \hht{(and the} optimizations it implements\hht{)}, and (ii)~it provides a
scalable approach \hht{to} testing \hht{for} \rh bit flips. Indeed, since different manufacturers deploy very
different TRR solutions as we show in Section~\ref{sec:rev-dram}, trying to obtain a detailed understanding of the
full behavior of every TRR-protected memory module is not practical. Even
so, we will demonstrate that these details in most cases do not \hht{get in
the way of finding} effective patterns: \n was able to automatically find access patterns
that trigger bit flips on modules we did \emph{not} analyze, and even on
mobile platforms \hht{using} \lpddr{} \hht{chips}---albeit in a simplified
\hht{way}.

\subsection{Design}
\label{sec:bb-test:design}

Based on the observations in Section~\ref{sec:rev-dram}, \n' fuzzing strategy
is based on two parameters: \emph{Cardinality} and \emph{Location}.

\vspace{2mm}
\noindent\textbf{Cardinality.} Cardinality represents the number of
aggressor rows hammered. We show in Section~\ref{sec:rev-dram:hnx} that
some \hht{modules} require a large number of aggressor rows to overflow the sampler
and induce \hhc{bit flips}. For instance, Figure~\ref{fig:smg:std-refi}
indicates that we need at least 7 rows to observe \hht{bit flips} \hht{in
module} \smg{15}. On the other hand, increasing the cardinality too much is
counterproductive. In particular, a \hhn{DRAM module} cannot carry out more than a
certain number of activations within the $64\,ms$ interval between two
refreshes of the same row. \hhn{The maximum number of row activations that
can be performed within $64\,ms$} mainly depends \hht{on} the \emph{row cycle time} ($t_{RC}$) that defines
the number of clock cycles between two \activate commands to the same bank.
In most modules $t_{RC} \approx 45\,ns$. It follows that the maximum number
of activations that we can perform within a $64 ms$ interval is
$1.4\times10^{6}$ ($64\,ms \div 45\,ns$). \hhn{If we tune} the fuzzer to hammer each
aggressor row at least 50K times (see Section~\ref{sec:rev-dram:hnx}), the
upper limit for the cardinality \hhn{is} 28 rows.

\vspace{2mm}
\noindent\textbf{Location.} Based on the results of
Section~\ref{sec:rev-dram:samsung}, we know that the sampler may \hhn{have
dependence} on row addresses. Thus, we want to randomize the location of the
aggressors to maximize the probability of bypassing address-dependent TRR mitigations.
Moreover, by picking the access pattern randomly, we implicitly
\hht{bypass} any feature of the sampler in the time domain. That is,
regardless of the \hht{design} of the sampler (\hhf{command-order}-based or frequency-based),
choosing random values for \hhn{the distances} between the aggressors
also randomizes \hht{the aggressors'} relative positions in the access pattern. Given a set
of aggressors, we choose to activate them in a round-robin fashion since
our experiments show that other strategies do not bring benefits in terms
of \hhf{the} number of \hht{bit flips}. 

\vspace{3mm}
\noindent\textbf{Fuzzing strategy.} \n evaluates \hhn{randomly-generated} access
patterns based on the number of unique \hht{bit flips}. It generates the patterns by
randomizing the cardinality and location parameters. If a bank contains $n$
rows, evaluating the combinations of all $n$ rows taking $k$ at a time ($k < n$)
would be impractical as $n$ is in the order of tens of thousands in modern
DRAMs. To obtain results within a reasonable \hhn{time frame}, the fuzzer therefore
allocates a smaller chunk of memory, spanning a subset of rows \hht{across} different
banks, and builds \rh{} \hht{access} patterns that respect the geometry of the memory
configuration~\cite{pessl2016drama}. The number of patterns that the fuzzer can
test in a given time frame is determined by the number of hammering rounds
(i.e., activations $\div$ cardinality). We pick this value such that we generate
activations that cover more than $3 \times \textit{\text{refresh period}}$. This
configuration makes sure that the victim rows are hammered for at least an
entire $64\,ms$ interval before their refresh.

%
%

\subsection{TRRespass-ing over DDR4}
\label{sec:bb-test:eval}
We evaluate our fuzzer and all other experiments on an Intel Core i7-7700K,
mounted on an ASUS STRIX Z270G motherboard. We acquire a set of \nDimms memory
modules produced by the three leading DRAM manufacturers (currently holding
around 95\% of the market~\cite{dram-share}). As shown in Table~\ref{tbl:dimms},
the set consists of 16 modules from vendor \smg{}, 12 from \mcr{}, and 14 from \hnx{}. 
We tested all the memory modules
singularly to draw conclusions about the individual chips. 
We ran \n for more than 6
hours on each module, scanning a memory chunk of 128 adjacent rows \eeh{from each
bank}. 
We now describe the results obtained through \n' black-box analysis.

\vspace{2mm}
\noindent\textbf{\Manysided \rh.} In one of our initial tests, \n assembled
a very simple and elegant access pattern that turned out to be effective on
most \mcr{} modules: \textit{assisted double-sided}. That is, a
double-sided pattern with a ``sidekick'' row. As shown in
Figure~\ref{fig:assisted-double-sided}, \hhn{this} pattern hammers rows $x-1,
x+1$, similarly to double-sided \rh, plus an extra one ($x+n$, where
$n>2$). 

\begin{figure}[!hb]
\centering\hfill
\begin{subfigure}[t]{.3\linewidth}
	\centering\includegraphics[width=1\linewidth]{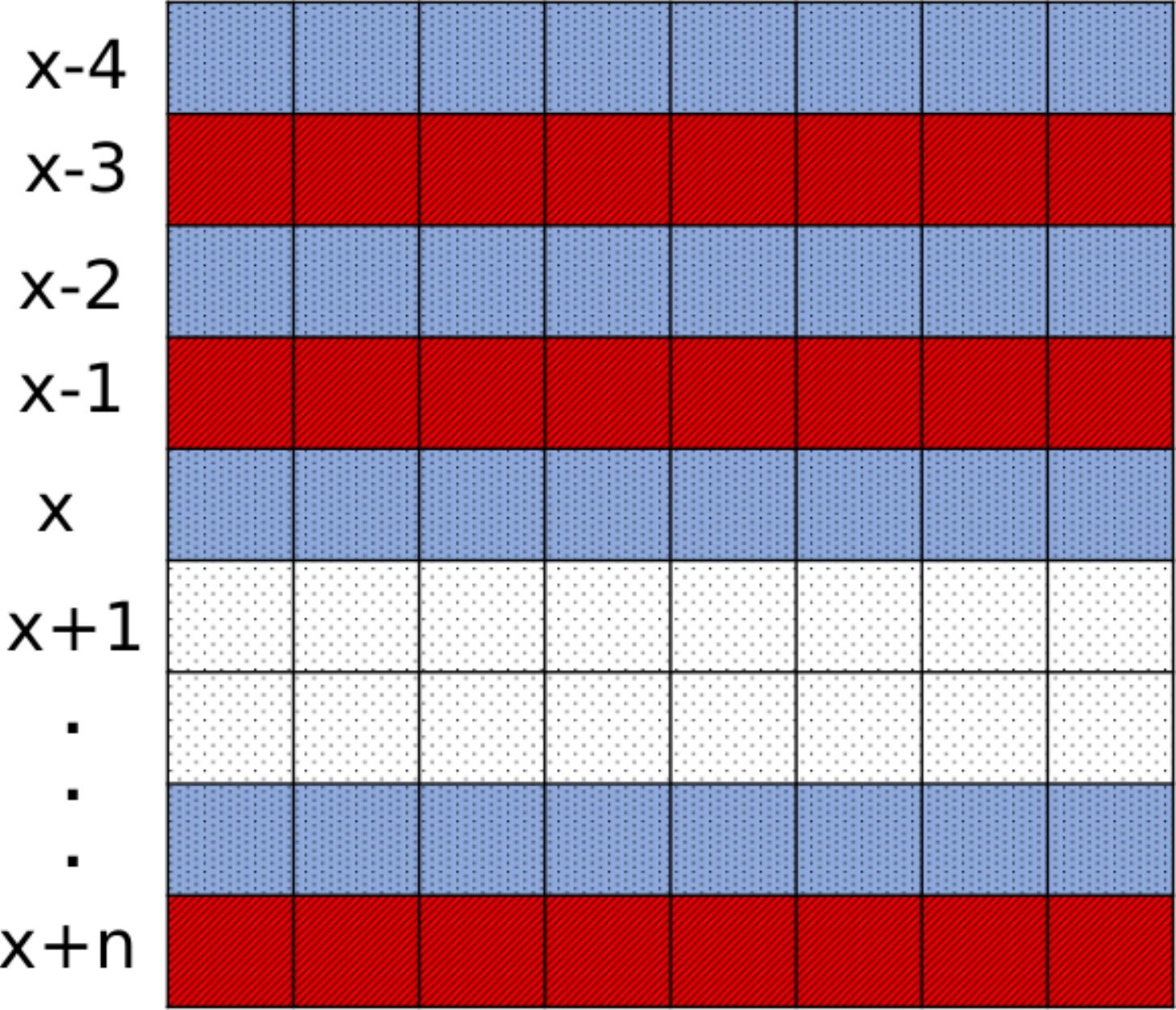}
    \captionsetup{width=1.3\linewidth}
    \vspace{-5.8mm}
    \caption{Assisted double-sided} \label{fig:assisted-double-sided}
\end{subfigure} \hfill
\begin{subfigure}[t]{.3\linewidth}
	\centering\includegraphics[width=1\linewidth]{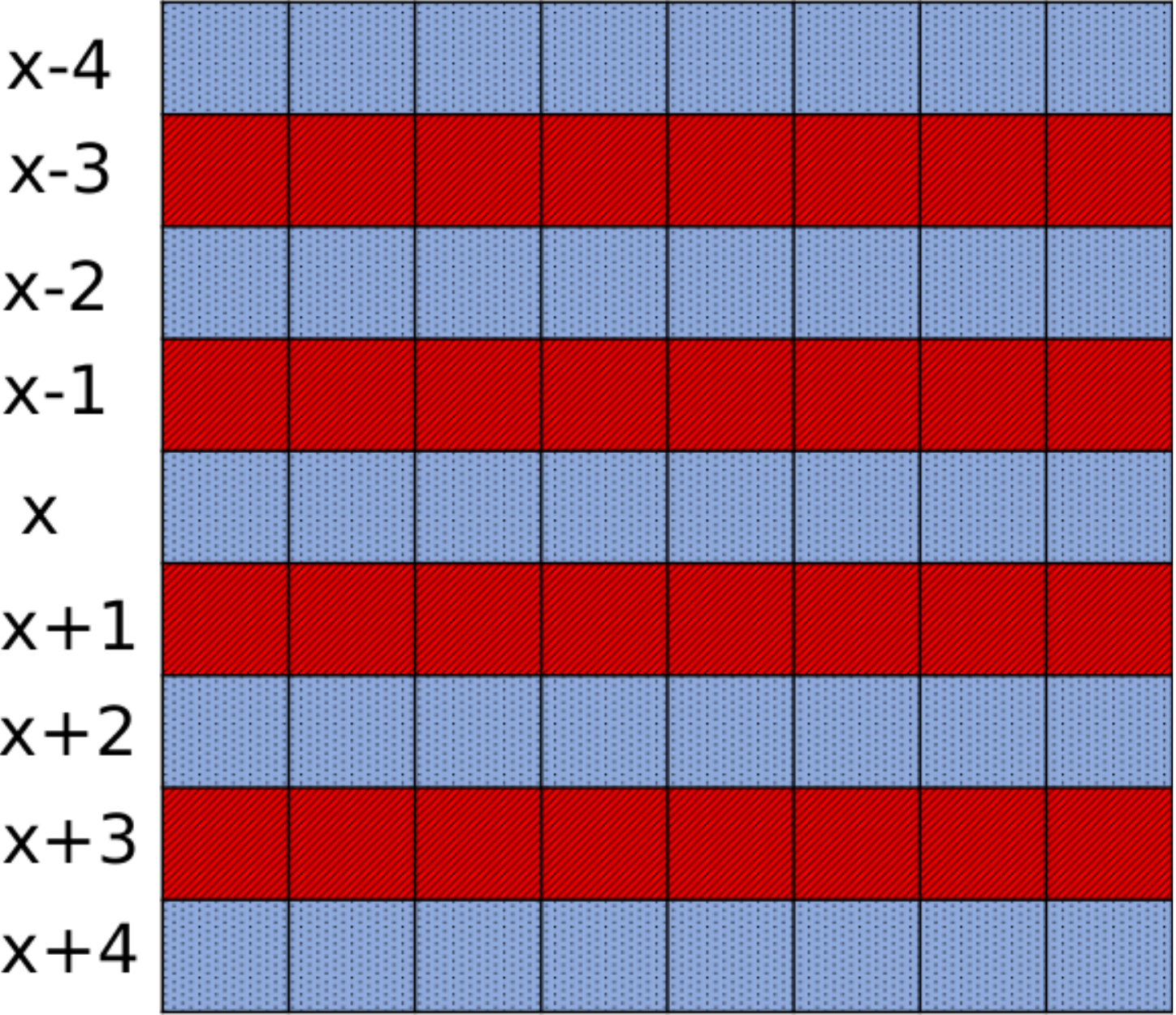}
	\caption{4-sided} \label{fig:quad-sided}
\end{subfigure} \hfill\hbox{}
    \caption{Hammering patterns \hhn{dis}covered by \n. Aggressor rows are in red (\protect\tikz \fill[red] (0.1,0.1) rectangle (0.3,0.3);) and victim rows are in blue (\protect\tikz \fill[blue] (0.1,0.1) rectangle (0.3,0.3);).}
\label{fig:new-variants}
\end{figure}

The analysis on all the \nDimms{} \hhn{DRAM modules} then allowed us to generalize the
assisted double-sided access pattern to a broader class of access patterns
which we call \emph{\Manysided \rh}. \hhn{Our} results show
that an attacker can benefit from \hht{sophisticated} hammering patterns that
exploit repeated accesses to \emph{many} aggressor rows. We now refer to the
discovered patterns using the nomenclature \nsided{$n$} where $n$ is the
\emph{cardinality} of the pattern. For instance, assisted double-sided
which is effective on \mcr{} \hhn{modules}
(Figure~\ref{fig:assisted-double-sided}), falls under the category of
\nsided{3} \rh.  Note that while we omit the \emph{location} of the
aggressors from this discussion, this parameter in some \hht{cases} does play
a role in the effectiveness of the pattern and we further discuss it in
Appendix~\ref{apdx:patterns}.

\vspace{3mm}
\noindent\textbf{Results.} \n{} \hhn{dis}covered effective access patterns for
\nVulnDimms of the \nDimms TRR-protected memory modules in our set.
Table~\ref{tbl:dimms} \hhn{reports} the results for the number of
access patterns identified and the \hht{structure} of the most effective pattern.
One interesting insight we gain from our analysis is that \emph{there is
not a single effective access pattern per module}. In fact, we can see that
all the modules where \n induces bit flips are vulnerable to at least two
different access patterns. On \mcr{} modules, we \hht{could} identify
access patterns on 4 out of the 12 modules we analyzed, and always
with simple \nsided{4} and \nsided{3} patterns as presented in
Figure~\ref{fig:new-variants}. On the other hand, none of these patterns
appear to work on the other vendors' chips. For example, in
Figure~\ref{fig:pairs-vs-flips}, we show the number of \hhn{aggressor rows} required
to trigger bit flips on \hht{module} \smg{10}. \hht{We can see that
no \hht{bit flip} can be triggered with fewer than 8 aggressor rows}. \n
successfully triggers bit flips on 7 of 16 \smg{} modules,
with several very different patterns. \smg{4} and \smg{15} are mainly vulnerable to the
\nsided{9} variant\hhn{,} \smg{10} and \smg{11} to different variants of the
\nsided{10} pattern, 
\hhn{and \smg{8} and \smg{9} to a \nsided{19} pattern.}
\pp{On \hnx{} modules, \n discovers effective
\hht{\rh} patterns \hhn{on only} \nHynixVuln of \nHynixTot modules.}
\hhn{We observe that \hnx{12} and \hnx{13} are vulnerable to
\nsided{9} and \nsided{10} hammering patterns.}\footnote{\pp{\hhn{While} \n identifies effective
\hht{\rh} access patterns only on \nVulnDimms out of \nDimms modules, this does not
mean that the other modules are immune to \rh. Similarly, these results do
not necessarily show that that memory modules from a specific vendor are
\hhc{more or less} vulnerable than \hhn{modules from other vendors}. Similar to regular software fuzzers,
it may simply be a matter of time and better strategies to find \hht{access}
patterns \hht{that lead to bit flips}.} \hhn{Our testing is also
\hhc{\emph{not}} exhaustive due to limited testing time and resources.}}



\begin{figure}[!ht]
	\centering
	\includegraphics[width=0.9\linewidth]{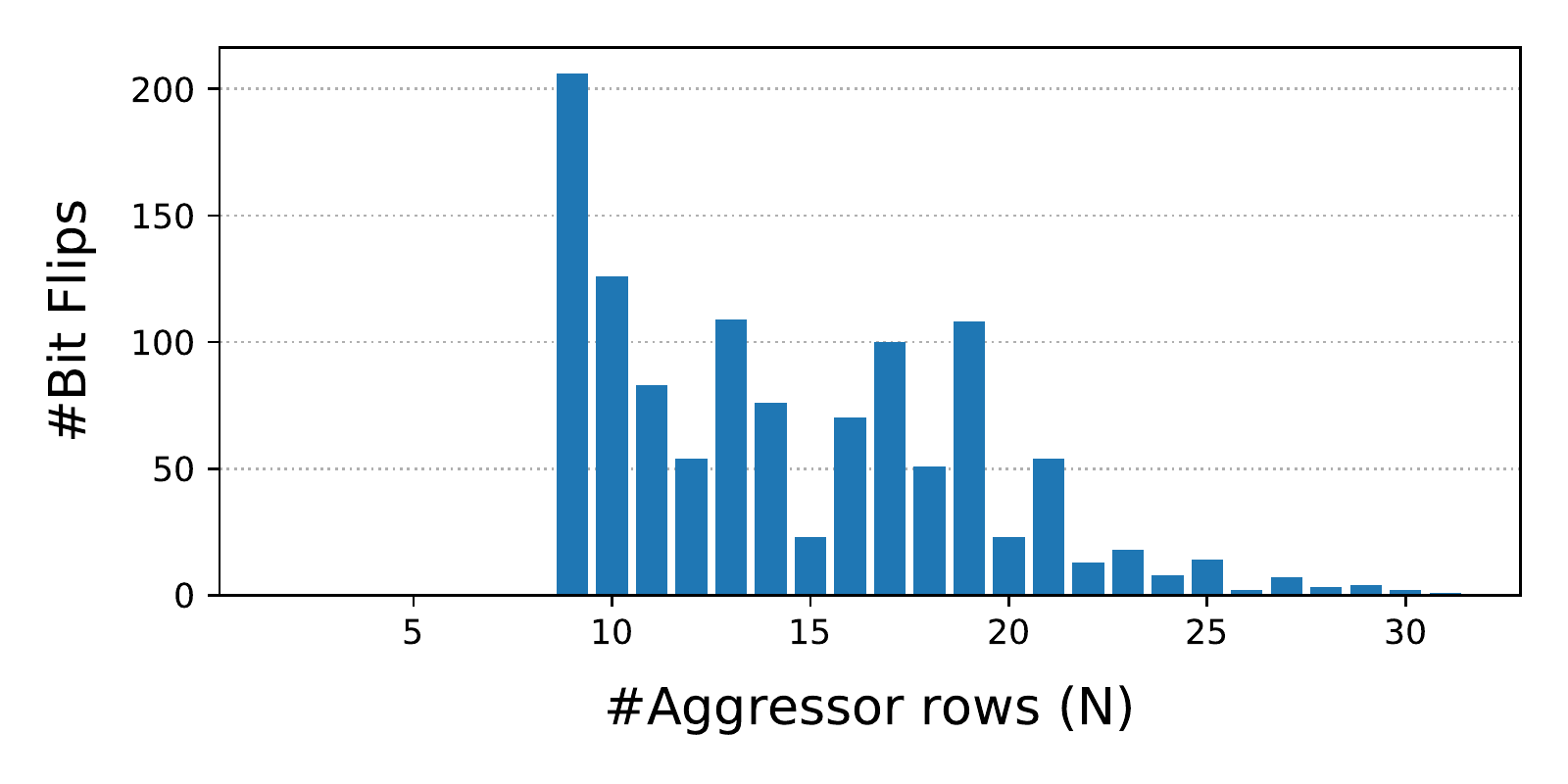}
    \caption{\textbf{Bit flips \pp{vs. \hhn{number of aggressor rows}}.}
    \pp{Module \smg{10}: Number of bit flips triggered with \emph{N-sided}
    \rh for varying number of \emph{N} on Intel Core i7-7700K. Each
    aggressor row is one row away from \hhn{the} closest \hhn{aggressor row} (i.e.,
    \texttt{VAVAVA}... configuration) and \hhn{aggressor rows} are hammered in a
    round-robin fashion.}} \label{fig:pairs-vs-flips}
\end{figure}


\vspace{3mm}
\noindent\textbf{A scalable framework.} The results of \n on \hht{module} \smg{15}
demonstrate how a black-box approach can be extremely beneficial. In
Section~\ref{sec:rev-dram}, we describe how complex it can be to reproduce
the optimal access pattern \hht{discovered using SoftMC on a CPU system}.
\hht{In contrast}, \n discovers two very successful access patterns that generate a
\hht{significant number} of bit flips automatically. 



\subsection{TRRespass on \lpddr}
\label{sec:bb-test:lpddr4}
In order to understand how widespread the issue is, we implement a
simplified version of \n for \aarch to test LPDDR4~\cite{jedec2014lpddr4} and
LPDDR4X~\cite{jedec2017lpddr4x} \hht{chips} on mobile phones. Due to the fragmented nature
of the Android ecosystem and the limited privileges (and resources)
available on some of these devices, we drop one of the two fundamental
parameters used in our previous design: \emph{location}. In other words,
we simply map a big chunk of memory and find a pool of addresses that
belong to the same bank~\cite{pessl2016drama}. Van der Veen et
al.~\cite{vanderveen2016drammer, vanderveen2018guardion} rely on uncached
memory due to the lack of cache flushing instructions on ARMv7. This
restriction does not apply any longer on \aarch{}\hhn{, and thus we do not
use uncached memory}. \hht{In our experiments,}
\n{} \hhn{dis}covers effective hammering patterns on \nVulnPhones of the
\nPhones devices, proving that \hht{TRR-protected} mobile platforms are
\hht{also still \hhn{vulnerable} to \rh} (\hhn{data is shown in} Table~\ref{tbl:lpddr4}). \hht{Not all
mobile platforms} report information about the memory
manufacturer \hht{and we do not} have fine grained control over the memory
allocations. As a result, we \hht{cannot} draw any conclusion with regard
to the extent of the vulnerability on \lpddr. Furthermore, phones from the
same model can use DRAM chips from different manufacturers. This means
that even if \n finds \rh bit flips on a certain phone from a specific
model, another phone from the same model may not exhibit these bit flips.
Similarly, the opposite can also be true.

\begin{table}[!t]
\newcommand\dgs{$^{\dagger}$}
\newcommand\empt{$^{\ }$}
\newcommand\nt{\texttt{N/A}}
\pit{check the manufacturers}
\centering
\renewcommand{\arraystretch}{1.2}
    \caption{\textbf{\lpddr results.} \hhn{Mobile} \hhc{phones} tested against \n on \aarch
    sorted by production date. We found \hht{bit flip inducing \rh}
    patterns on \nVulnPhones out of \nPhones \hhn{mobile} phones.}
\begin{threeparttable}
    \resizebox{\linewidth}{!}{
\begin{tabular}{ l c c c c}
\toprule
    \emph{\makecell{\hhn{Mobile}\\\hhc{Phone}}} & \emph{Year} & \emph{SoC} & \emph{\makecell{Memory\\(GB)}} &
    \emph{\makecell{\hhn{Found}\\Patterns}}\\ 
\midrule

Google Pixel                        & 2016   & MSM8996         & 4\tnote{$\dagger$}         & \yes  \\
\stripe
Google Pixel 2                      & 2017   & MSM8998         & 4                          & \notavail  \\
\makecell[l]{Samsung\\G960F/DS}        & 2018   & \makecell{Exynos\\9810}     & 4           & \notavail  \\
\stripe
Huawei P20 DS                       & 2018   & Kirin 970       & 4                          & \notavail  \\
Sony XZ3                            & 2018   & SDM845          & 4                          & \notavail  \\
\stripe
HTC U12+                            & 2018   & SDM845          & 6                          & \notavail  \\
LG G7 ThinQ                         & 2018   & SDM845          & 4\tnote{$\dagger$}         & \yes  \\
\stripe
Google Pixel 3                      & 2018   & SDM845          & 4\                         & \yes  \\
Google Pixel 4                      & 2019   & SM8150          & 6\                         & \notavail  \\
\stripe
OnePlus 7                           & 2019   & SM8150          & 8\                         & \yes  \\
\makecell[l]{Samsung\\G970F/DS}        & 2019   & \makecell{Exynos\\9820}     & 6\          & \yes  \\
\stripe
Huawei P30 DS                       & 2019   & Kirin 980       & 6\                         & \notavail  \\
\makecell[l]{Xiaomi Redmi\\Note 8 Pro} & 2019   & \makecell{Helio\\G90T}      & 6\          & \notavail  \\

\bottomrule
\end{tabular}
} 
\begin{tablenotes}[online]
\scriptsize
\item[$\dagger$] LPDDR4 (not LPDDR4X)
\end{tablenotes}
\end{threeparttable}
\label{tbl:lpddr4}
\vspace{-4mm}
\end{table}





\section{Evaluation}
\label{sec:eval}
In this section, we \hhn{systematically} evaluate \hht{our} \nDimms{} \hht{DDR4 DRAM} modules against the
optimal \rh access pattern (i.e., the one that yields the most \hht{bit flips})
\hht{identified} by \n for each module.

\subsection{Results}
\label{sec:eval:ddr4}

We test \hht{each of the} \nDimms modules \hhn{using the
most-bit-flip-incurring \rh pattern that
we discover for each module in Section~\ref{sec:bb-test:eval}.}
\pp{For every module, we
perform a sweep over 256\,MB of contiguous physical
memory}.\footnote{\hhn{We avoid testing the entire capacity of the DRAM
modules and instead test 256\,MB of each module to reduce testing time.
We note that this could potentially cause \n to miss the most \rh-prone
portions of a module. \hhc{Thus, we believe \n is likely to be more
effective than what we report in this paper.}}} We then examine the memory for
\rh{} \hht{bit flips} in both \hh{\emph{true}} cells and \emph{anti}
cells~\cite{kim2014flipping, liu2013experimental}. In
other words, we look for both $1\to0$ and $0\to1$ bit flips. We \hh{show} the
results for all the \nDimms modules in Table~\ref{tbl:dimms}. We \hhn{now}
provide a detailed explanation of these results by discussing them
\hht{separately for}  each DRAM vendor.\\

\vspace{3mm}
\noindent\textbf{Vendor \smg{}.} In Section~\ref{sec:bb-test}, we show
\hhn{\n can bypass} how mitigations from manufacturer \smg{}. \hh{We can} recover multiple effective access patterns for \tocheck{7 of the
16} modules in our \hh{experiments}. In Table~\ref{tbl:dimms}, we
\hh{provide} the number of \hht{bit flips} that we
observe on the vulnerable \smg{} modules. The results are worrisome: we
\hh{find} more than 16K \hht{bit flips} on average \hhn{across the 7
vulnerable modules}. 
\hht{In addition to the \hhn{large} number of bit flips, we also observe
that the bit flips occur with significantly fewer row activations on
\hhn{vendor \smg{}'s}
DDR4 modules compared to previous generation \hhn{DDR3} DRAM devices. For example, on
\smg{8} and \smg{9}, we \hhn{can effectively} perform 19-sided \rh{}
\hhn{with as few as}
\tildesmall{}45K row activations to each of the effective
aggressor rows (i.e., the aggressor rows adjacent to the target victim
row(s)) within \hhn{the} 64ms refresh period. \hhn{In contrast}, Kim et
al.~\cite{kim2014flipping} show that bit flips occur with \tildesmall{}139K or more
DRAM row activations on \hhn{older} DDR3 modules.}\\



\vspace{-1mm}
\noindent\textbf{Vendor \mcr{}.} In Section~\ref{sec:bb-test}, we describe assisted
double-sided (i.e., \nsided{3}) and \nsided{4} \hhn{hammering} as two effective patterns against
a subset of our memory modules from vendor \mcr{}. However, the \hht{low
bit flip counts} in
Table~\ref{tbl:dimms} show that \hhn{bypassing} the \hh{TRR} mitigation on
these modules is non-trivial. \tocheck{We \hh{run} further \hh{experiments}
on these modules to understand \hhn{the} limited number of \hht{bit flips}
\hhn{we observe}}.
We \hh{make} two \hh{observations. 
\hhn{First, when we repeat for multiple iterations the same \rh experiment using the
aggressor rows that we know can cause bit flips, we observe a varying
number of bit flips in the victim row(s) across different iterations.}
Figure~\ref{fig:M0-rounds} shows the number of \hh{bit flips} that we can
trigger on a specific row, \hhf{using} $3$ aggressors in \hh{module} \mcr{0}.
\hht{We observe from the figure that different iterations \hhf{(i.e.,
samples)} of the same test reveal a different number of bit flips in the same victim row}.
\hh{Second, when} hammering the same module in a multi-DIMM configuration
(i.e., two identical modules on the same system), we often observe more
\hht{bit flips}. These results hint at the presence of a parameter \n
cannot (yet) bypass. \hhn{The} \hh{fact that we occasionally observe a} large
number of \hht{bit flips} suggests \hh{that} these modules are quite
susceptible to \rh, and \hh{causing more bit flips} may be only a matter of
improving our fuzzing strategy. \\ }


\begin{figure}[!ht]
	\centering
	\vspace{-2mm}
	\includegraphics[width=0.9\linewidth]{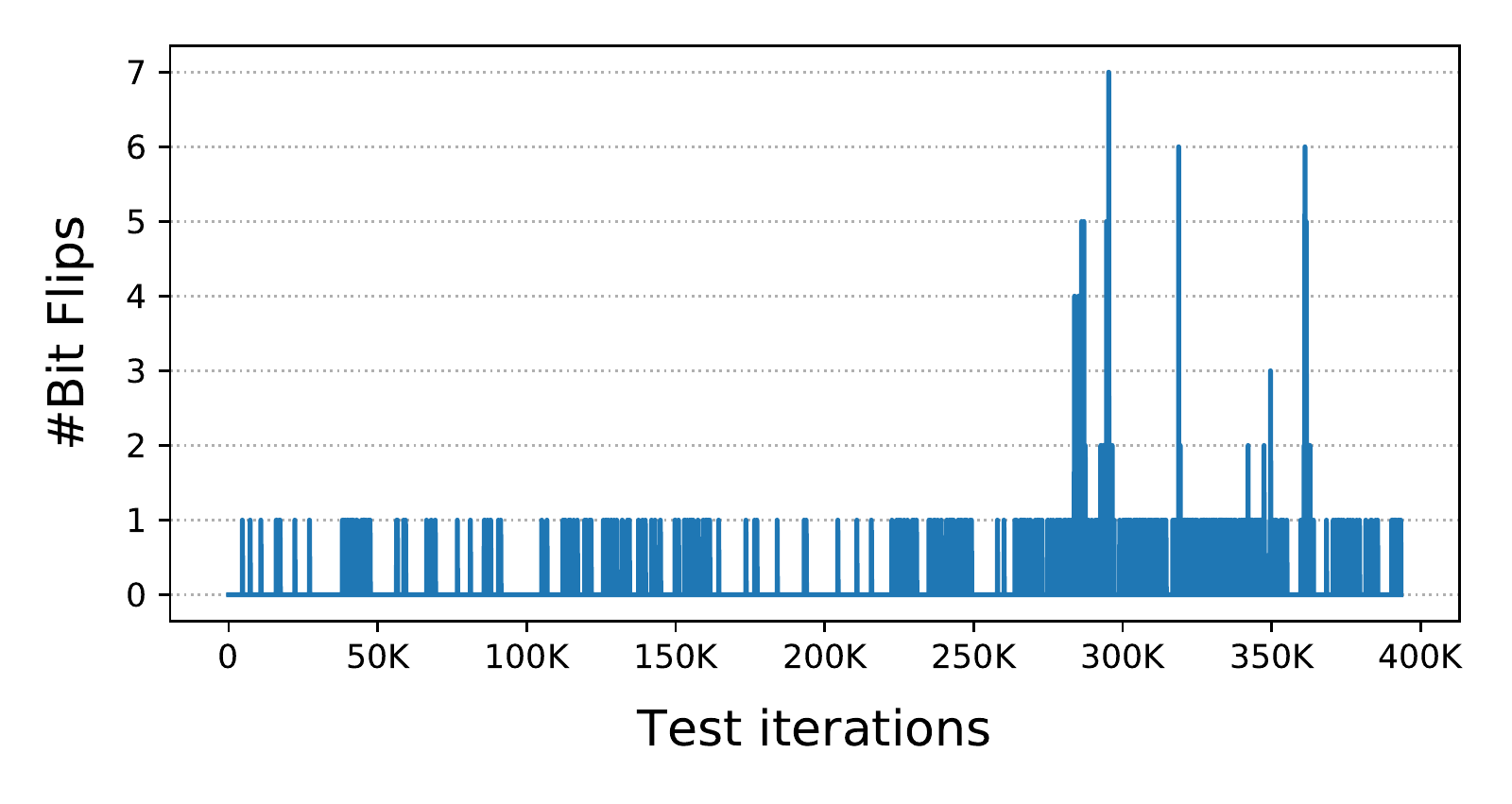}
	\vspace{-1mm}
    \caption{\textbf{Bit flips vs. \hhn{Test iterations}}. Module
    \mcr{0}: Number of bit flips over \hhn{test iterations}. In each
    \hhn{iteration}, the aggressor rows are
    hammered for \hht{three} refresh intervals.}
	\label{fig:M0-rounds}
\end{figure}


\vspace{-1mm}
\noindent\textbf{Vendor \hnx{}.}
\pp{\n identifies effective patterns on \nHynixVuln of \nHynixTot \hnx{}
modules. However, we see a steep drop in the number of bit flips on modules
from newer generations. \hhn{Module} \hnx{12}, produced before 2015, is the
oldest and most vulnerable module in our test set (Table~\ref{tbl:dimms}).
\hhn{Modules} of newer generations are less vulnerable (if at all) to the
patterns identified by \n. This suggests that the in-DRAM TRR
implementation has evolved over time.} \pp{\hht{We perform further
experiments} on \hnx{13} \hht{to confirm} this hypothesis. We
\hht{discover} that, instead of performing a single targeted refresh
\hht{during each regular} refresh operation, the TRR mitigation employed by
\hnx{13} \hht{performs} \hhn{\emph{multiple}} targeted refreshes
\hht{during each regular refresh operation}. While we can confirm that
\hht{recent} DRAM \hht{chips are} still vulnerable \hht{to \rh}, further
research is required to better understand \hhn{newer} TRR mitigations
\hhn{and} to find more effective fuzzing strategies \hhn{against them}.}

\subsection{\hht{Increasing the Refresh Rate}}
\label{sec:eval:half-refi}
As mentioned in Section~\ref{sec:rev-dram:preliminaries}, the memory
controller issues a \refresh command to the memory device every
$7.8\,\mu{}s$, to ensure that \hhn{all} cells are refreshed within a $64\,ms$
interval. Doubling \hht{(or even quadrupling)} the refresh rate \hht{(i.e.,
double-refresh)} was proposed in the
past~\cite{intel2014ptrr, aweke2016anvil, refresh-lenovo, kim2014flipping,
AppleRefInc} as \hht{an immediate}
countermeasure against \rh attacks, since \hht{doing so reduces} the amount of
time required for hammering.  As discussed in Section~\ref{sec:rev-mc}, some
server platforms employ double-refresh as default behavior or enable it
when a non-TRR-compliant \hhn{DRAM module} \hht{is} in use. This is usually not the case
on consumer platforms.\footnote{\pp{As we report in Table~\ref{tbl:dimms},
we occasionally detect double-refresh behavior on \hht{particular} DRAM
\hht{modules}. This suggests \hht{that} the memory controller may employ
module-dependent \hhn{mechanisms for \rh mitigation}.}} However, \trefi can (sometimes) be set in
the BIOS.
Although double-refresh was demonstrated in the past to not fully
\hhn{eliminate the \rh vulnerability}~\cite{aweke2016anvil,kim2014flipping}\hhc{,} the introduction of \hht{in-DRAM} TRR
may have changed the situation. In fact, since \hht{TRR} acts mainly
at refresh time, doubling the refresh operations could improve \hht{TRR's} security
guarantees, enforcing the inhibitor operations \hhn{more frequently}. To test this hypothesis, we
evaluate our modules against \n when running \hht{them} with double refresh.

\hhn{Our} experiment reveals the presence of new hammering patterns that are still
able to trigger bit flips in \hht{three modules} (\hhn{as indicated in the
rightmost column in} Table~\ref{tbl:dimms}). 
This result further \hht{undermines} the efficacy of double refresh as a stopgap solution
against \rh even when in-DRAM \hht{TRR is} deployed.

\subsection{\hht{Repeatability of the Bit Flips}}
\label{sec:eval:reproducibility}
\pit{still have to look from here on}
Repeatability is a fundamental factor in \rh exploitation. The ability
\hht{to} reliably trigger a bit flip
\hht{repeatedly}~\hhn{\cite{kim2014flipping}} is what made \rh so
popular in adversarial scenarios\hhn{~\cite{razavi2016flip, frigo2018grand,
gruss2016rowhammer, bosman2016dedup, vanderveen2016drammer, xiao2016one,
tatar2018throwhammer, seaborn2015exploiting-1, mutlu2019rowhammer,
mutlu2017rowhammer}}.

We study the repeatability of these \hht{many-sided \rh} bit flips to better understand their properties.
We pick one DRAM module per DRAM vendor (\smg{14}, \mcr{1}, \hnx{13}) and we run the best pattern for each\pit{do we need to test the \hnx{12}?}
\hht{module}. When \hhn{a bit flip} occurs, we try to \hht{repeat} it.
\hhn{Our} experiment confirms
that bit flips are repeatable in a reliable way for all the modules. 
However, it may require multiple attempts before obtaining the
same \hht{bit flip} \hhn{again} and sometimes we may observe many other spurious
\hht{bit flips} generated by the same pattern (\hhn{see our analysis of}
\hhf{Vendor} \mcr{} \hhn{modules} in
Section~\ref{sec:eval:ddr4}). We discuss the implications \hht{of this
phenomenon} for the
exploitation \hht{of these bit flips} in Section~\ref{sec:exploitation}.  



\section{Exploitation with \n}
\label{sec:exploitation}

\n generates \manysided \rh patterns to bypass TRR on modern DDR4 modules.
While such access patterns are more sophisticated than standard \rh{} \hht{access} patterns~\cite{kim2014flipping, gruss2018another,
tatar2018defeating}, we now show \hhn{that} their practical exploitability is not only
possible, but also similar, in spirit, to existing state-of-the-art \rh
attacks. For this purpose, we show how we craft \hht{many-sided \rh}
exploits using the general \rh exploitation framework used by prior work in
the area~\cite{razavi2016flip}. The exploitability investigated by the
framework revolves around three fundamental steps: 
\begin{enumerate*}[label=(\roman*)]
\item \emph{Memory templating}, 
\item \emph{Memory massaging}, and 
\item \emph{Exploitation}.
\end{enumerate*}\\

\vspace{-2mm}
\noindent\textbf{Memory Templating.} In this step, the attacker scans memory
with \rh access patterns, looking for vulnerable memory pages (or
\emph{templates}) where one or more bits can be flipped at a specific offset.
For templating to be successful, an attacker needs to \hhn{use} the desired
patterns when accessing DRAM. Prior work has already demonstrated the
feasibility of \hhn{using} double-sided \rh patterns using either huge (2MB)
pages~\cite{razavi2016flip} or a variety of side channels to identify physically
contiguous memory
ranges~\cite{frigo2018grand,vanderveen2016drammer,rambleed,spoiler}. For
\manysided \rh, we can use the former mechanism as long as we can fit all the
aggressor rows in a single huge page (similar to double-sided \rh).
\tocheck{This is possible for simple variants such as \nsided{3} \rh, but not for
    complex variants such as \nsided{19} (which \hht{may} require two or more consecutive
	huge pages). However, many of these modules vulnerable to lengthy patterns
    are \hht{also} vulnerable to a series of different other patterns (often shorter).
    \hht{Moreover}, the results on \lpddr, where we simply hammer random
	addresses belonging to the same bank, demonstrate that the location of the
	aggressors is not always a fundamental parameter---relaxing the
	assumptions for the attacker. In the case where only extended patterns
    (e.g., \nsided{19}) are effective or in the absence of huge pages
    \hhn{in} the system, we can still use a \hht{variety} of page allocator
    side channels~\cite{frigo2018grand,vanderveen2016drammer,rambleed} or
    speculative side channels~\cite{spoiler, ridl} to locate a sufficiently
    large contiguous memory chunk to fit our \manysided \rh patterns and
    template memory.\\ }

\vspace{-2mm}
\noindent\textbf{Memory Massaging.} Once vulnerable templates are available, the attacker needs to implement
some form of memory massaging to lure the victim into mapping the target
data onto one of the available templates. Any of the memory
massaging techniques described in prior work still apply with no modifications
to \manysided \rh, given that memory massaging is
pattern-agnostic~\cite{frigo2018grand, vanderveen2016drammer,
gruss2016rowhammer, gruss2018another, bosman2016dedup, razavi2016flip,
vanderveen2018guardion}.\\

\vspace{-2mm}
\noindent\textbf{Exploitation.} Once the target data is mapped onto the target
template, the attacker needs to trigger the same \rh bit flips using the
previously templated access patterns to complete the final exploitation
step. For this step to be successful, the attacker needs to ensure that,
with high probability, (i)~the templated bit flips are \hht{repeatable}, and
(ii)~there are no spurious (non-templated) bit flips in the victim page.
Prior work has shown that these assumptions hold in practice for
state-of-the-art attacks based on standard access patterns. Compared to
such patterns, \manysided patterns incur similar (albeit lower)
\hht{repeatability}, as discussed in Section~\ref{sec:eval:reproducibility}. In
practice, this means the attacker may have to \hht{perform} the access patterns
multiple times for reliable exploitation. Moreover, to ensure there are no
spurious bit flips across runs, the attacker can trivially mask irrelevant
columns in the aggressor rows as shown in previous work~\cite{rambleed,
cojocar2019eccploit, 236248} or otherwise use these bit flips as part of a
compatible attack vector (e.g., corrupting multiple bits of a cryptographic
key~\cite{razavi2016flip}).

Overall, \n-based exploitation is very similar to existing \rh attacks. As
shown in the next section, once effective \manysided access patterns are
available, an attacker can reliably mount real-world \rh attacks on modern
DDR4 systems in a matter of minutes.

\subsection{Exploitation on DDR4}
\label{sec:eval:epxloitation}
Armed with (repeatable) templates, we now study the effectiveness of
different \rh exploits on modern DDR4 systems. To this end, we implement
three \hht{example} attacks: (i) the original \rh exploit targeting PTEs
(\emph{Page Table Entries}) to obtain kernel privileges from Seaborn and
Dullien~\cite{seaborn2015exploiting-1}, (ii) the RSA exploit from Razavi et
al.~\cite{razavi2016flip} \hhn{that corrupts} public keys to gain access to a
co-hosted VM, and (iii) the \emph{opcode flipping} exploit on the
\texttt{sudo} binary from Gruss et al.~\cite{gruss2018another}. The PTE
exploit~\cite{seaborn2015exploiting-1} takes advantage of bit flips on the
\emph{Page Frame Number} (PFN) to probabilistically redirect the virtual to
physical mapping of an attacker-controlled page to another page table page.
This relies on page table spraying to increase the probability of
referencing another page table page with the corrupted PFN. The exploit
from Gruss et al.~\cite{gruss2018another} shows that it is possible to
target code pages in the page cache to compromise opcodes and bypass
permission checks on the \texttt{sudo} binary. Gruss et
al.~\cite{gruss2018another} report 29 vulnerable opcodes to use for this
purpose. Razavi et al.~\cite{razavi2016flip} propose an attack to
compromise an RSA public key stored in the page cache. They prove that
\hhn{causing a bit flip in the \emph{modulus} of \hhc{a} 1024-bit
\hhc{or} 2048-bit RSA public key makes the modulus} factorizable with a probability
of 12-22\%. For our analysis, we target a
2048-bit RSA public key.

\vspace{3mm} We assume an attacker capable of performing memory
massaging---placing an exploitable target on one of the vulnerable memory
pages---using any well-known technique~\cite{frigo2018grand,
vanderveen2016drammer, gruss2016rowhammer, gruss2018another,
bosman2016dedup, razavi2016flip, vanderveen2018guardion}.
Table~\ref{tbl:expl} presents our results for two sample \hhn{modules} for each vendor---the most and least vulnerable from the same manufacturer. As part of
our analysis, we also record {\large$\uptau$} (i.e., time to template
\hht{a single} row),
since many-sided \rh requires more time to carry out templating
compared to previous \rh variants. As expected, we see a \hht{large} discrepancy across the different modules, which matches the \hht{largely} different number of \hht{bit
flips} reported in Table~\ref{tbl:dimms}. In the case of \mcr{} modules, where \n is able to generate very few bit flips, we are unable to reproduce any
attack. On the other hand, on the other 4 \hhn{modules} from vendors \smg{} and
\hnx{} we can (overall) find \hhn{templates} to reproduce all the attacks. On
\hnx{12}\hhn{,} we can reproduce the PTE attack~\cite{seaborn2015exploiting-1} in as little as 2.3\,$s$, while the RSA-2048 exploit~\cite{razavi2016flip}, when successful, can take up to 39\,$m$\,48\,$s$ (\smg{4}). Bypassing \texttt{sudo} permission checks~\cite{gruss2018another} turned out to be possible only on \hnx{12} in 54\,$m$\,16\,$s$.
Note that we assume existing templating strategies as is: we did
not attempt to craft more sophisticated attacks, since our goal is solely
to \hht{test} existing \rh variants.
Overall, our results show that \rh still \hht{presents} a significant threat to the security of modern DDR4 systems\hht{, even in the presence of in-DRAM TRR mitigations}.

\begin{table}[!h]
\centering
\renewcommand{\arraystretch}{1.2}
\caption{\textbf{Time to exploit.} Time to find the first exploitable
    template on \pp{two sample modules from each DRAM vendor}.}
\begin{threeparttable}
    \resizebox{\linewidth}{!}{
\begin{tabular}{ l l m{0.1pt} c c c}
\toprule
    \emph{Module} & {\large$\uptau$}\,\emph{(ms)} && \emph{\hht{PTE}}\cite{seaborn2015exploiting-1} & \emph{RSA-2048}\cite{razavi2016flip} & \emph{\texttt{sudo}}\cite{gruss2018another}\\ 
\midrule
\stripe
\smg{14} & 188.7 		&& 4.9s 			& 6m\,27s 			& \notavail 	\\
\smg{4}  & 180.8		&& 38.8s 			& 39m\,28s 			& \notavail 	\\
\stripe
\mcr{1}  & 360.7 		&& \notavail 		& \notavail 		& \notavail 	\\
\mcr{2}  & 331.2 		&& \notavail 		& \notavail 		& \notavail 	\\
\stripe
\hnx{12} & 300.0		&& 2.3s				& 74.6s				& 54m16s 		\\
\hnx{13} & 180.9		&& 3h\,15m			& \notavail			& \notavail 	\\		

\bottomrule
\end{tabular}
} 

    \vspace{1mm}

    \scriptsize
{$\uptau$}:\ Time to template a \hht{single} row: time to
    fill the \hht{victim and aggressor} rows $+$ hammer time $+$  time to
    scan the row. \hfill \mbox{} 
\end{threeparttable}
\label{tbl:expl}
\end{table}


\setstretch{.88}

\section{Related Work}
\label{sec:rel-work}

\noindent\textbf{\rh.} In their seminal work, Kim et
al.~\cite{kim2014flipping} are the first to \hhd{rigorously introduce and} characterize the \rh
vulnerability. Following this work, a \hhd{large number of} of attacks compromising a variety of different
systems~\cite{seaborn2015exploiting-1, gruss2016rowhammer,
vanderveen2016drammer, frigo2018grand,  xiao2016one, razavi2016flip,
vanderveen2018guardion, tatar2018throwhammer, bhattacharya2016curious,
bhattacharya2018advanced, carre2018openssl, fournaris2017exploiting,
jang2017sgx, poddebniak2018attacking, aga2017good, zhang2018triggering} and
characterization studies\hhd{~\cite{cojocar2019eccploit, tatar2018defeating,
qiao2016new,gruss2018another, cojocar2020are}} emerged\hhd{, as described
in a recent retrospective article~\cite{mutlu2019rowhammer}}. Prior \hhd{works} rely on three main
classes of \rh patterns to induce bit flips: (i) single-sided, (ii)
double-sided, and (iii) one-location \rh. None of these techniques are
effective against modern DDR4 modules with in-DRAM \rh mitigations.
Lanteigne~\cite{thirdio_nodate_rowhammer,thirdio_2016_rowhammerAMD}
\hhd{proposes} a technique (\hhd{called} regional \rh), where a small 2\,MB region
(e.g., \hhd{a} Linux hugepage) is hammered using multiple software threads to increase the
\eeh{DRAM row activation rate. In fact, we are not the first to use the term
    \nsided{n} \rh for $n=4$, as Lanteigne refers to his technique as
    \emph{quad-sided}~\cite{thirdio_2016_rowhammerAMD}. However, his technique does not provide a clear or
    methodical way of picking aggressor rows that are close to each other
    in a bank, and instead aims to maximize the number of row activations.}
     We show that merely maximizing the number of activations is not
     sufficient to bypass in-DRAM \rh mitigations.

\vspace{1mm} 
\noindent\textbf{Software-based mitigations.} Herath and
Fogh~\cite{herath2015these} and Aweke et al.~\cite{aweke2016anvil}
suggested ``hybrid'' mitigations based on hardware performance counters to
detect suspicious hammering-like activity. Other mitigations\hhn{,} such as
CATT~\cite{brasser2017catt} and
GuardION~\cite{vanderveen2018guardion}\hhn{,} try
to enforce DRAM-based data isolation to prevent \rh attacks from corrupting
sensitive data. Nevertheless, recent work has shown how these mitigations
cannot stop more sophisticated attacks~\cite{gruss2018another,
frigo2018grand}. With the correct DRAM mapping functions,
ZebRAM~\cite{konoth2018zebram} can protect the entire system by extending
isolation to the entire DRAM. Unfortunately, ZebRAM becomes expensive when the
active working set of an application is larger than half of DRAM capacity.

\vspace{1mm} 
\noindent\textbf{Hardware-based mitigations.} Although
doubling the refresh rate or using ECC memory \hht{are
immediately-deployable solutions, they have proven} insufficient to stop
\rh~\cite{aweke2016anvil, cojocar2019eccploit, kim2014flipping}. \hhf{Other
hardware-based mitigation techniques have been proposed~\cite{lee2019twice,
son2017making, seyedzadeh2017cbt} but, to our
knowledge, these \hhn{have} not been deployed in real systems.} 
\hhc{Kim et al. propose Probabilistic Adjacent Row Activation
(PARA)~\cite{kim2014flipping}, which is a low overhead mechanism to prevent
\rh bit flips. When a row is activated, with a very small probability, PARA
refreshes rows adjacent to
the activated row. A variant of PARA,
Hardware RHP, appears to be employed by some Intel memory
controllers~\cite{tq2020hw-rhp, versa2019hw-rhp, intel2017hw-rhp, omron2019hw-rhp}. This is a new
\rh measure in the memory controller and its robustness is yet to be
independently validated.}
In recent years, TRR has
become \hht{the} hardware-based \rh mitigation of choice, first \pph{deployed in the MC} on DDR3 systems
and then \pph{in-DRAM} on DDR4. 
While DDR3 systems have been widely studied,
only a few studies have reported \rh bit flips on
DDR4~\cite{lipp2018nethammer, thirdio_nodate_rowhammer, gruss2018another}.
Compared to our analysis, such studies have induced bit flips on selected earlier-generation DDR4 modules.
In contrast, we study several generations of DDR4 modules (including the
\hh{most-recent off-the-shelf devices}) and find that, while standard access patterns are no
longer effective, \hhn{new} \manysided \rh patterns can still induce bit flips on \hht{many
TRR-protected} DDR4 modules in the market today.

\setstretch{.86}
\section{Conclusion}
\label{sec:conclusion}

This paper \hht{shows} that, despite significant mitigation efforts,
\pp{modern DDR4 \hhn{DRAM} systems} are still vulnerable to \rh{}
\hhn{bit flips}---and even more vulnerable than \hhn{DDR3 DRAM systems}, once the
mitigations are bypassed. In particular, we demonstrate that
\emph{Target Row Refresh} (TRR), \hht{publicized} by \pp{CPU and DRAM} vendors
as the definitive solution to \rh, \hh{can be bypassed to cause \rh bit
flips}. First, we show
that TRR is an umbrella term for a variety of mitigations deployed at
the memory controller or in DRAM \hht{chips}. Second, we analyze common
TRR implementations in the memory controller (using timing side
channels) and \hht{in DRAM chips} (using \hht{an} FPGA-based memory
controller\hhn{, SoftMC}).
\hhn{Our analysis shows that the consumer CPUs we test rely on in-DRAM TRR
\hhc{to mitigate} the \rh vulnerability and do not employ TRR at the memory
controller level.}
\hhn{We} discover that modern (\hht{in-DRAM}) TRR implementations are generally
\hht{vulnerable to} \emph{\manysided \rh}, \hht{a new hammering strategy
that \hhn{hammers \emph{many} (i.e., at least 3)} aggressor rows
\hhn{concurrently}}. Finally, we
present \n, a black-box
\manysided \rh fuzzer that, unaware of the implementation of the
memory controller or the \hht{DRAM chip}, can still find sophisticated
\hh{hammering} patterns to mount real-world attacks for many of the
\hhn{DDR4} \hht{DRAM} modules
\hht{in} the market. Our results provide evidence that the pursuit of
effective \rh mitigations must continue and that the \emph{security by
obscurity} strategy of DRAM vendors puts computing systems at risk for extended
periods of time.


\section*{Disclosure}

We disclosed our new \rh attacks to all affected parties in November of 2019.
This triggered an industry-wide effort in addressing the issues raised in this paper.
Unfortunately, due to the nature of these vulnerabilities, it will take time before effective mitigations will be in place. 
Further developments on these vulnerabilities are tracked under CVE-2020-10255.
The paper remained confidential until the public disclosure date of March 10, 2020.

\section*{Acknowledgments}

We would like to thank the anonymous reviewers for their valuable feedback and Robin Webbers for helping us in our analysis of \lpddr systems. 
This work was supported by the European Union's Horizon 2020 research and
innovation programme under grant agreements No. 786669 (ReAct) and No. 825377
(UNICORE), by Intel Corporation through the Side Channel Vulnerability ISRA, and
by the Netherlands Organisation for Scientific Research through grants NWO
639.023.309 VICI ``Dowsing'', NWO 639.021.753 VENI ``PantaRhei'', and NWO
016.Veni.192.262. This paper reflects only the authors' view. The funding
agencies are not responsible for any use that may be made of the information it
contains.

\setstretch{1}

\bibliographystyle{IEEEtranS}
\bibliography{references}

\appendices


\section{TRR-compliant memory}
\label{apdx:trr-compliant}

In Section~\ref{sec:rev-mc}, we define TRR-compliant memory. Here we expand
on \hh{this} concept, also explaining the difference between TRR-compliant and
pTRR-compliant memory.

The \mac field is a field of one byte located at byte 41 on the SPD of a DDR3
module~\cite{jedec2014spdDDR3} and byte 7 on \hht{the SPD of a} DDR4
module~\cite{jedec2015spdDDR4}. This field reports information about the
\hht{module's} resiliency to \rh. In the single byte allocated to the \mac value
inside the SPD~\cite{jedec2015spdDDR4, jedec2014spdDDR3}, only the 6 least
significant bits are used to store information about the \hht{module's} limits in
the form of \mac and \tmaw (Figure~\ref{fig:mac}), where \mac is the
\emph{Maximum Activate Count} and \tmaw is the \emph{Maximum Activate
Window}, which simply acts as a multiplier for \mac (Figure~\ref{fig:mac}).
The remaining \hhn{two} most significant bits are flagged as reserved.  As we
mention in Section~\ref{sec:rev-mc} the \mac value can take \hht{three}
configurations:

\begin{itemize}
    \item \emph{unlimited}\hhn{,} as value \texttt{0b1000}; 
\item \emph{untested}, as value \texttt{0b0000};  or
\item discrete values from \emph{200K} to \emph{700K} with steppings of $+100K$---values \texttt{0b0001} to \texttt{0b0110}.
\end{itemize}


\begin{figure}[!h]        
	\centering
    \vspace{-3mm}
	\includegraphics[width=0.8\linewidth]{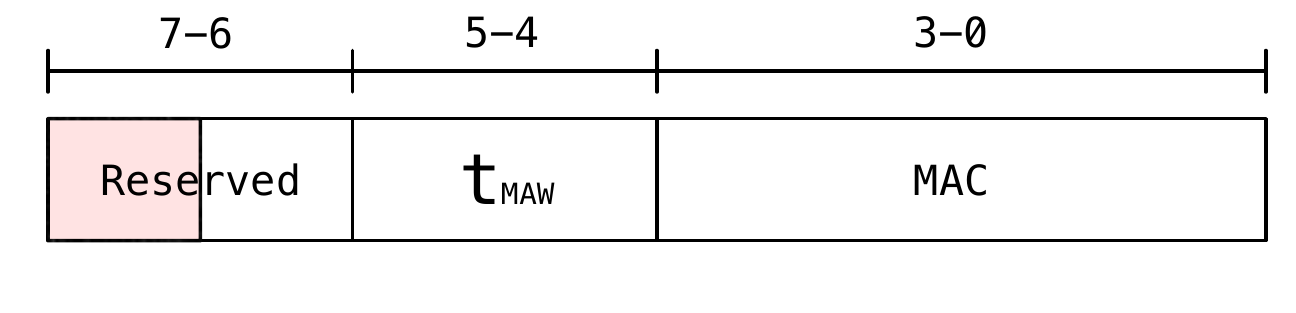}
    \vspace{-3mm}
    \caption{\textbf{SPD's \mac field.} Bit 7 needs to be \hhn{set} in order to enable pTRR~\cite{intel2014ptrr}.}
	\label{fig:mac}
    \vspace{-3mm}
\end{figure}

In one of our early experiments, we discovered that our definition of
TRR-compliant modules slightly diverges from Intel's definition of
pTRR-compliant modules~\cite{intel2014ptrr}. In fact, we discovered that
in order to enable pTRR, bit 7 (one of the reserved bits) needs to be set.
If not, regardless of the \mac and \tmaw values, the system treats the module
as non-compliant. This is likely a legacy feature which stems from the fact
that pTRR~\cite{intel2014ptrr} was introduced before TRR became part of the
JEDEC standard~\cite{jedec2014spdDDR3}.

\section{\n-ing patterns}
\label{apdx:patterns}


\hhn{In Section~\ref{sec:bb-test:eval}, we explain the new
\nsided{$n$} hammering patterns we use in our experiments. We now provide a
more general definition of these hammering patterns.}

\hhn{\n randomizes two parameters: \emph{cardinality} and \emph{distance}.}
\hhn{Cardinality} and distance \hhn{together define} a novel hammering
pattern \hhn{that} we refer to as \dnsided{$d$}{$n$}
\rh. 
\hhn{The pattern consists of $\frac{n}{2}$ \emph{pairs} of aggressor rows, where
the two aggressor rows in each pair are placed one victim row apart
(similar to double-sided \rh). The
distance $d$ defines the number of rows between the aggressor row pairs.
For example, the \dnsided{$3$}{$4$} pattern contains two aggressor row
pairs (four aggressor rows in total), and the two aggressor row pairs are three rows
apart \hhc{from each other}. The \nsided{n} pattern, which we refer to throughout the paper, is
another example\hhc{,} where the distance
between the aggressor row pairs is one row.}
\ee{Figure~\ref{fig:nsided_distance} shows the number of bit flips
\hhn{that occur in}
\hhn{module} \smg{10} when \hhn{we use}
the \dnsided{D}{10} hammering pattern \hhc{while} \hhn{sweeping} the parameter
$D$. We \hhn{note} that the number of bit flips \hhn{increases and
decreases as we vary $D$}, reaching
its maximum \hhn{at} $D=12$. This observation confirms that the distance
\hhn{between aggressor row pairs} has a primary role in
assembling an effective hammering pattern.}


\begin{figure}[!h]
	\centering
    \vspace{-2mm}
	\includegraphics[width=0.8\linewidth]{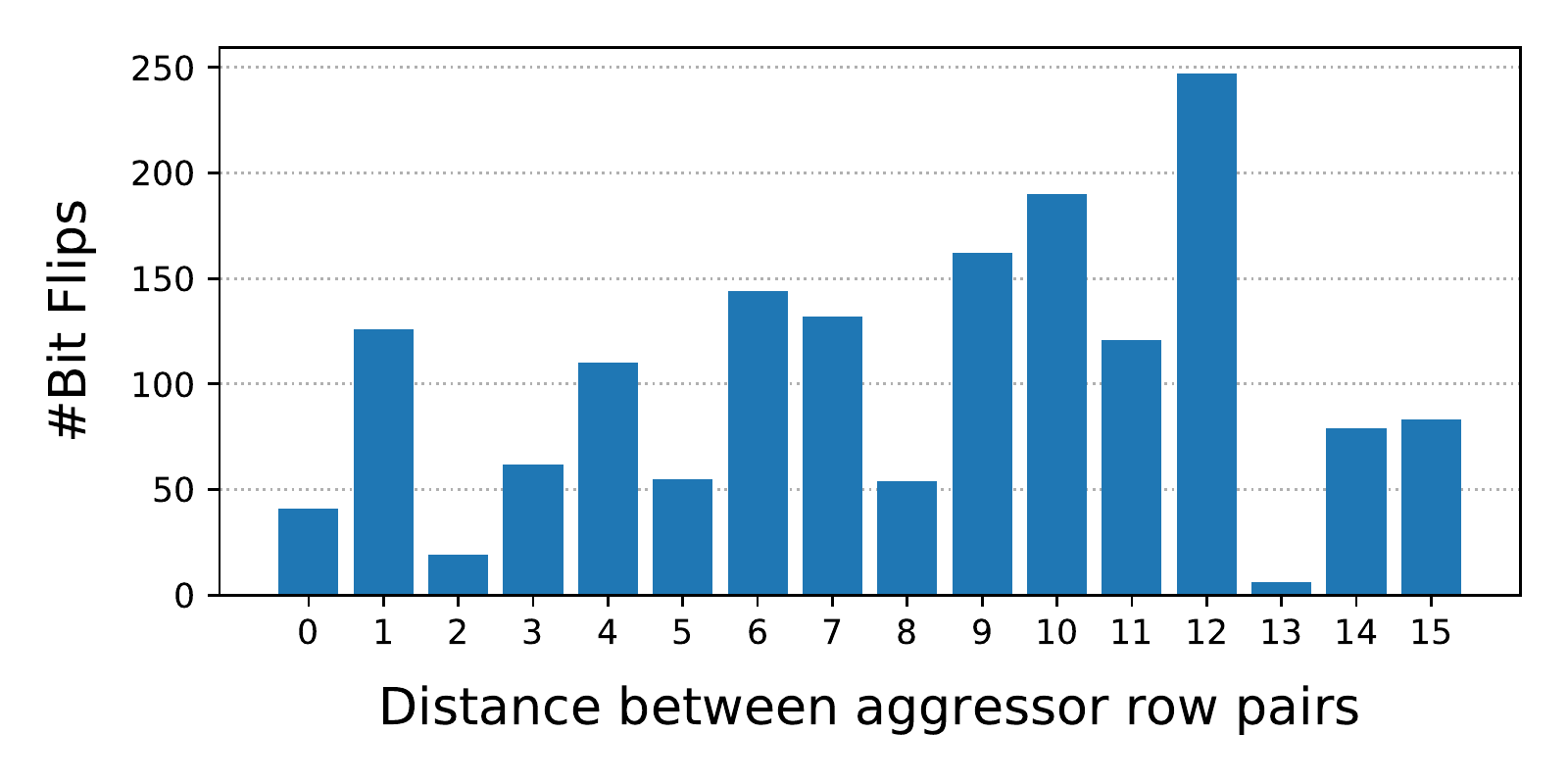}
    \vspace{-2mm}
    \caption{\textbf{Bit flips induced by \dnsided{D}{10} \rh \hhn{pattern
    as a function of} D}. X-axis \hhc{plots} the
    distance between each aggressor \hh{row pair}. Y-axis reports the number
of unique bit flips.}
	\label{fig:nsided_distance}
\end{figure}


\end{document}